\documentclass[preprint]{aastex}

\usepackage{graphicx}
\usepackage{natbib}

\def\sn{iPTF\,13ajg}
\def\lsim{\mathrel{\hbox{\rlap{\lower.55ex \hbox {$\sim$}}\kern-.0em
\raise.4ex \hbox{$<$}}}} 
\def\gsim{\mathrel{\hbox{\rlap{\lower.55ex \hbox {$\sim$}}\kern-.0em
\raise.4ex \hbox{$>$}}}} 
\def\lya{Ly$\alpha$}
\def\ha{H$\alpha$}

\def\l{$\lambda$}
\def\kms{km s$^{-1}$}
\def\1star{$^{\star}$}
\def\2star{$^{\star\star}$}
\def\3star{$^{\star\star\star}$}
\def\4star{$^{\star\star\star\star}$}

\newcommand{\Msun}{\mbox{M$_\odot$}}

\newcommand{\Msunyr}{\mbox{M$_\odot$ yr$^{-1}$}}

\newcommand{\cm}{\mbox{cm$^{-2}$}}

\newcommand{\ergscm}{\mbox{erg s$^{-1}$} cm$^{-2}$}

\newcommand{\ergscmA}{\mbox{erg s$^{-1}$ cm$^{-2}$ \AA$^{-1}$}}

\newcommand{\persec}{s$^{-1}$}

\begin{document}


\title{The hydrogen-poor superluminous supernova \sn\ and its host
  galaxy in absorption and emission}




\author{
  Paul M. Vreeswijk\altaffilmark{1,2},
  Sandra Savaglio\altaffilmark{3,4,5},
  Avishay Gal-Yam\altaffilmark{1},
  Annalisa De Cia\altaffilmark{1},
  Robert M. Quimby\altaffilmark{6,7},
  Mark Sullivan\altaffilmark{8},
  S. Bradley Cenko\altaffilmark{9,10,11},
  Daniel A. Perley\altaffilmark{12,13},
  Alexei V. Filippenko\altaffilmark{9},
  Kelsey I. Clubb\altaffilmark{9},
  Francesco Taddia\altaffilmark{14},
  Jesper Sollerman\altaffilmark{14},
  Giorgos Leloudas\altaffilmark{14,15},
  Iair Arcavi\altaffilmark{16,17},
  Adam Rubin\altaffilmark{1},
  Mansi M. Kasliwal\altaffilmark{18},
  Yi Cao\altaffilmark{12},
  Ofer Yaron\altaffilmark{1},
  David Tal\altaffilmark{1},
  Eran O. Ofek\altaffilmark{1},
  John Capone\altaffilmark{19},
  Alexander S. Kutyrev\altaffilmark{20},
  Vicki Toy\altaffilmark{19},
  Peter E. Nugent\altaffilmark{21,9},
  Russ Laher\altaffilmark{22},  
  Jason Surace\altaffilmark{22}, and
  Shrinivas R. Kulkarni\altaffilmark{12}
}

\altaffiltext{1}{Department of Particle Physics and Astrophysics,
  Weizmann Institute of Science, Rehovot 7610001, Israel}

\altaffiltext{2}{Benoziyo Fellow; email: paul.vreeswijk@weizmann.ac.il}

\altaffiltext{3}{Max Planck Institute for Extraterrestrial Physics,
  85748 Garching bei M\"unchen, Germany}

\altaffiltext{4}{European Southern Observatory, 85748 Garching bei
  M\"unchen, Germany}

\altaffiltext{5}{Physics Department, University of Calabria, 87036
  Arcavacata di Rende, Italy}

\altaffiltext{6}{
  Kavli Institute for the Physics and Mathematics of the Universe (WPI),
  Todai Institutes for Advanced Study, The University of Tokyo
  5-1-5 Kashiwanoha, Kashiwa-shi, Chiba, 277-8583, Japan
}

\altaffiltext{7}{Department of Astronomy, San Diego State University,
  San Diego, CA 92182, USA}

\altaffiltext{8}{School of Physics and Astronomy, University of
  Southampton, Southampton SO17 1BJ, UK}

\altaffiltext{9}{Department of Astronomy, University of California,
  Berkeley, CA 94720-3411, USA}

\altaffiltext{10}{Astrophysics Science Division, NASA Goddard Space
  Flight Center, Mail Code 661, Greenbelt, MD, 20771, USA}

\altaffiltext{11}{Joint Space-Science Institute, University of
  Maryland, College Park, MD 20742, USA}

\altaffiltext{12}{Astronomy Department, California Institute of
  Technology, MC 249-17, 1200 East California Blvd, Pasadena CA 91125,
  USA}

\altaffiltext{13}{Hubble Fellow}

\altaffiltext{14}{Department of Astronomy, The Oskar Klein Center,
  Stockholm University, AlbaNova, 10691 Stockholm}

\altaffiltext{15}{Dark Cosmology Centre, Niels Bohr Institute,
  Copenhagen University, Juliane Maries Vej 30, 2100 Copenhagen O,
  Denmark}

\altaffiltext{16}{Las Cumbres Observatory Global Telescope Network,
  6740 Cortona Dr., Suite 102, Goleta, CA 93117, USA}

\altaffiltext{17}{Kavli Institute for Theoretical Physics, University
  of California, Santa Barbara, CA 93106, USA}

\altaffiltext{18}{The Observatories, Carnegie Institution for Science,
  813 Santa Barbara Street, Pasadena, CA 91101, USA}

\altaffiltext{19}{Department of Astronomy, University of Maryland,
  College Park, MD 20742, USA}

\altaffiltext{20}{Astrophysics Science Division, NASA Goddard Space
  Flight Center, Mail Code 665, Greenbelt, MD, 20771, USA}

\altaffiltext{21}{Computational Cosmology Center, Lawrence Berkeley
  National Laboratory, 1 Cyclotron Road, Berkeley, CA 94720, USA}

\altaffiltext{22}{Spitzer Science Center, MS 314-6, California
  Institute of Technology, Jet Propulsion Laboratory, Pasadena, CA
  91125, USA}

\begin{abstract}
  We present imaging and spectroscopy of a hydrogen-poor superluminous
  supernova (SLSN) discovered by the intermediate Palomar Transient
  Factory: \sn. At a redshift of $z$=0.7403, derived from narrow
  absorption lines, \sn\ peaked at an absolute magnitude $M_{u,{\rm
      AB}}=-22.5$, one of the most luminous supernovae to date. The
  $uBgRiz$ light curves, obtained with the P48, P60, NOT, DCT, and
  Keck telescopes, and the nine-epoch spectral sequence secured with
  the Keck and the VLT (covering 3 rest-frame months), are tied
  together photometrically to provide an estimate of the flux
  evolution as a function of time and wavelength, from which we also
  estimate the bolometric light curve. The observed bolometric peak
  luminosity of \sn\ is $3.2\times 10^{44}$~erg~\persec, while the
  estimated total radiated energy is $1.3\times10^{51}$~erg. We detect
  narrow absorption lines of \ion{Mg}{1}, \ion{Mg}{2}, and
  \ion{Fe}{2}, associated with the cold interstellar medium in the
  host galaxy, at two different epochs with X-shooter at the VLT, at a
  resolving power $R\approx6000$. From Voigt-profile fitting, we
  derive the column densities log $N$(\ion{Mg}{1}) $= 11.94\pm0.06$,
  log $N$(\ion{Mg}{2}) $=14.7\pm0.3$, and log $N$(\ion{Fe}{2})
  $=14.25\pm0.10$.  These column densities, as well as the \ion{Mg}{1}
  and \ion{Mg}{2} equivalent widths of a sample of hydrogen-poor SLSNe
  taken from the literature, are at the low end of those derived for
  gamma-ray bursts (GRBs), whose progenitors are also thought to be
  massive stars. This suggests that the environments of SLSNe and GRBs
  are different. From the nondetection of \ion{Fe}{2} fine-structure
  absorption lines, we derive a strict lower limit on the distance
  between the supernova and the narrow-line absorbing gas of 50~pc.
  The neutral gas responsible for the absorption in \sn\ exhibits a
  single narrow component with a low velocity width, $\Delta
  V=76$~\kms, indicating a low-mass host galaxy. No host-galaxy
  emission lines are detected, leading to an upper limit on the
  unobscured star-formation rate of SFR$_{\rm
    [O~II]}<0.07~\Msunyr$. Late-time imaging shows the host galaxy of
  \sn\ to be faint, with $g_{\rm AB}\approx27.0$ and $R_{\rm AB}\geq
  26.0$~mag, which roughly corresponds to $M_{B,{\rm Vega}} \gtrsim
  -17.7$~mag.
\end{abstract}



\section{Introduction}
\label{sec:introduction}

The advent of high-cadence imaging of large fractions of the sky, by
surveys such as the Texas Supernova Search
\citep[TSS;][]{2006PhDT........13Q}, the Catalina Real-Time Transient
Survey \citep[CRTS;][]{2009ApJ...696..870D}, the Palomar Transient
Factory \citep[PTF;][]{2009PASP..121.1334R,2009PASP..121.1395L}, and
Pan-STARRS \citep{2010SPIE.7733E..12K}, has led to the recognition
\citep{2011Natur.474..487Q} of a class of superluminous supernovae
(SLSNe), which includes previously unidentified transients such as
SCP~06F6 \citep{2009ApJ...690.1358B} and SN~2005ap
\citep{2007ApJ...668L..99Q}. SLSNe are typically defined as SNe
reaching absolute magnitudes $M<-21$ \citep{2012Sci...337..927G},
about 10--100 times brighter than normal SNe. Their observed redshift
range is $z=0.02$--3.9 \citep{2012Natur.491..228C}, with a median of
$z\approx0.3$. SLSNe are rare; their total rate at $z\approx0.2$
\citep[][]{2013MNRAS.431..912Q,2014arXiv1402.1631M} is roughly
$10^{-3}$ times that of core-collapse SNe at a similar redshift
\citep{2008A&A...479...49B}.  To date, about 50 SLSNe have been
reported in the literature.

The inferred energetics of SLSNe require processes that are different
from those in classical SNe, probably involving very massive stars,
but the physical mechanisms powering these explosions are very poorly
understood \citep[for a review, see][]{2012Sci...337..927G}. As with
normal SNe \citep[for a review, see][]{1997ARA&A..35..309F}, SLSNe can
be divided into hydrogen-rich (Type II) and hydrogen-poor classes
(Type I).  Type II SLSNe exhibit strong hydrogen features in their
spectra, probably produced by a thick hydrogen envelope surrounding
the explosion which complicates identification of their nature. This
envelope is thought to be responsible for producing the energetics
observed by capturing the kinetic energy of the explosion and
converting it into thermal radiation at a sufficiently large distance,
similar to the model that has been invoked for Type IIn SNe
\citep{1982ApJ...258..790C,1994ApJ...420..268C,2010ApJ...724.1396O,
  2011ApJ...729L...6C,2012ApJ...756L..22M,2013MNRAS.428.1020M}.  Type
I SLSNe lack hydrogen features and reach the highest peak luminosities
among SLSNe \citep[but see][]{2013arXiv1310.1311B}, with very blue
spectra and copious ultraviolet (UV) flux persisting for many
weeks. Late-time spectra of at least some Type I SLSNe show features
of Type Ic SNe \citep{2010ApJ...724L..16P,2011Natur.474..487Q}. The
currently discussed models for producing the energy in Type I SLSNe
are the interaction of the SN with circumstellar material
\citep[CSM;][]{2011ApJ...729L...6C} without displaying hydrogen
features
\citep{2011Natur.474..487Q,2010arXiv1009.4353B,2012ApJ...760..154C,2014ApJ...785...37B},
and the spin down of a highly magnetic, rapidly spinning neutron star,
\citep[a magnetar;][]{2010ApJ...717..245K,2010ApJ...719L.204W}.

Among the hydrogen-poor class, several SLSNe have very slowly declining
light curves which can be explained by the presence of several solar
masses of radioactive nickel ($^{56}$Ni).  \citet{2012Sci...337..927G}
suggests identifying these hydrogen-poor, slowly declining events as
a separate class: Type R (for radioactive). These explosions are
thought to require an extremely massive ($\gtrsim50$~\Msun) progenitor,
possibly leading to a pair-instability supernova
\citep{2009Natur.462..624G}, although this is debated
\citep[see][]{2010ApJ...717L..83M,2012MNRAS.426L..76D,2013Natur.502..346N}.

Imaging shows that SLSNe typically explode in dwarf galaxies, with
likely low metallicities \citep{2011ApJ...727...15N}. The host
galaxies of a sample of Type I SLSNe exhibit similarities to galaxies
nurturing gamma-ray bursts \citep[GRBs;][]{2013arXiv1311.0026L}.
Because of their high luminosity, SLSNe are in principle excellent probes
of galaxies having a large fraction of massive stars at high redshift
\citep{2011Natur.474..487Q,2012ApJ...755L..29B}, in a similar way as
GRB afterglows have been used for more than a decade. SLSNe do not
reach the extreme luminosities provided by GRB afterglows, but they
stay luminous for a much longer time (at least in the rest-frame
optical range), allowing for less time-restricted follow-up
observations. Moreover, GRBs typically require a gamma-ray satellite
such as {\it Swift} \citep{2004ApJ...611.1005G} to rapidly detect and
accurately localize their afterglows \citep[but
  see][]{2013ApJ...769..130C}, whereas SLSNe can be detected from the
ground.

Rest-frame near-UV spectra of different SLSNe have revealed narrow
absorption lines of \ion{Mg}{2} and \ion{Fe}{2}, enabling a precise
redshift measurement for the host galaxy in the absence of emission
lines \citep{2011Natur.474..487Q}. However, to date, these spectra
have been taken at low resolution ($R=\lambda/\Delta\lambda \lesssim
1000$), which does not allow for meaningful constraints on the column
density or the presence of multiple components at different velocities
of the absorbing gas. The combination of apparent brightness and
redshift of \sn\ permitted us to obtain the first set of
intermediate-resolution ($R\sim6000$) spectra of a SLSN.

This paper is organized as follows. In Section \ref{sec:observations} 
we provide a description of our extensive
photometric and spectroscopic observations of \sn, which is followed
by a discussion of the photometric (Sect.~\ref{sec:photometry}),
spectral (Sect.~\ref{sec:spectra}), and bolometric evolution
(Sect.~\ref{sec:Lbol}). The narrow resonance absorption features of
\ion{Mg}{1}, \ion{Mg}{2}, and \ion{Fe}{2} that we detect in our
intermediate-resolution X-shooter spectra are analyzed in
Sect.~\ref{sec:narrowabslines}, while the absence of \ion{Fe}{2}
fine-structure lines is used alongside excitation modeling in
Sect.~\ref{sec:excitation} to derive a lower limit on the distance of
the narrow-line absorbing gas to \sn. The properties of the supernova
host galaxy are inferred in Sect.~\ref{sec:host}. We discuss our
results in Sect.~\ref{sec:discussion} and we briefly conclude with
Sect.~\ref{sec:conclusions}.

Unless noted otherwise, the uncertainties listed in this paper are at the
1$\sigma$ confidence level, while the limits are reported at
3$\sigma$. We adopt a luminosity distance to \sn\ at $z=0.7403$ of
$d_{\rm L}=4.6$~Gpc (or distance modulus $\mu=43.32$ mag), assuming
H$_0=70$~km~s$^{-1}$~Mpc$^{-1}$, $\Omega_{\rm m}=0.28$, and
$\Omega_{\Lambda}=0.72$ \citep{2013ApJS..208...19H}.  We note that the
cosmological parameters as derived by the Planck collaboration
\citep[H$_0=67.3$~km~s$^{-1}$~Mpc$^{-1}$, $\Omega_{\rm m}=0.315$,
$\Omega_{\Lambda}=0.685$;][]{2013arXiv1303.5076P} would result in
a luminosity distance that is 2.2\% larger than the one we adopt,
leading to absolute magnitude estimates that are 0.05 mag
brighter than those presented in this paper.

\section{Identification, observations, and data reduction}
\label{sec:observations}

\begin{figure}[h!]
  \centering
  \includegraphics[width=\hsize]{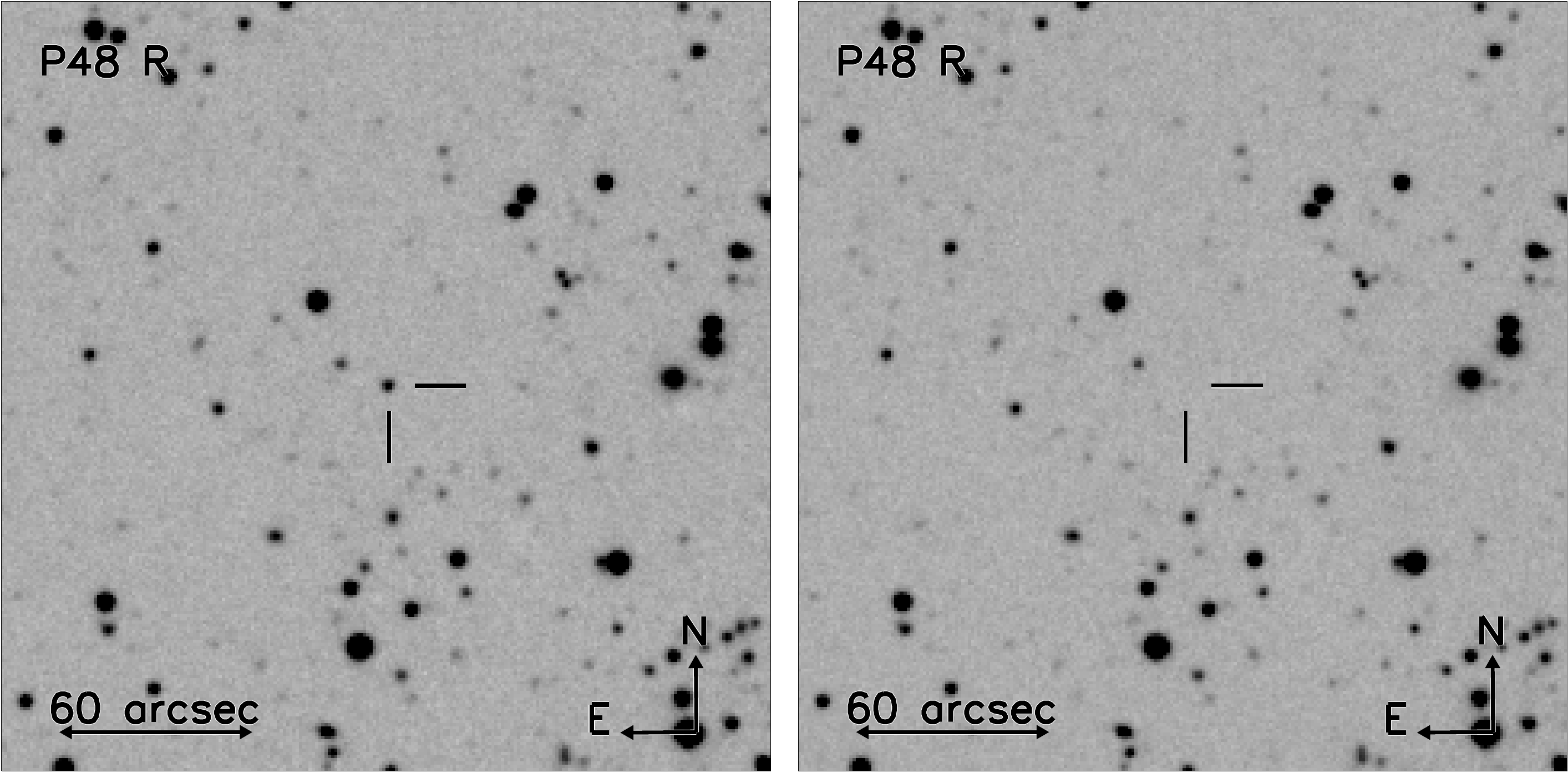}
  \caption{A combination of several Palomar 48~inch $R$-band images
    around the time of peak brightness of \sn\ (at left) and from
    before the explosion (at right); the latter image has an
    approximate limiting magnitude of
    $R=23.2$. \label{fig:images13ajg}}
\end{figure}

\sn\ was flagged as a transient source as part of the regular
operations of the intermediate Palomar Transient Factory \citep[iPTF;
  see][]{2009PASP..121.1334R,2009PASP..121.1395L} on 2013 April 7 (UTC 
dates are used throughout this paper). A low-resolution Keck/LRIS 
spectrum taken the next night allowed the transient 
to be classified as a Type I SLSN at $z=0.7403$. This redshift is
based on the detection of narrow absorption lines of \ion{Mg}{1},
\ion{Mg}{2}, and \ion{Fe}{2}. The sky coordinates of \sn\ are 
$\alpha = 16^{\rm h}39^{\rm m}03.95^{\rm s}$, $\delta = +37^\circ 01' 38.4''$ 
(J2000.0), with an uncertainty of 0.1\arcsec. 
At this location the Galactic extinction is estimated to
be low, $A_V=0.04$~mag \citep{2011ApJ...737..103S}. The field of
\sn\ is shown in Figure~\ref{fig:images13ajg}, with and without the
supernova present.

\sn\ was imaged with the Palomar 48~inch (P48) Oschin iPTF survey
telescope equipped with a 12k $\times$ 8k CCD mosaic camera
\citep{2008SPIE.7014E..4YR} in the Mould $R$ filter, the Palomar
60~inch and CCD camera \citep{2006PASP..118.1396C} in Johnson $B$ and
SDSS $gri$, the 2.56-m Nordic Optical Telescope (on La Palma, Canary
Islands) with the Andalucia Faint Object Spectrograph and Camera
(ALFOSC) in SDSS $ugriz$, the 4.3-m Discovery Channel Telescope (at
Lowell Observatory, Arizona) with the Large Monolithic Imager (LMI) in
SDSS $r$, and with the Low Resolution Imaging Spectrograph
\citep[LRIS;][]{1995PASP..107..375O} and the Multi-Object Spectrometer
for Infrared Exploration \citep[MOSFIRE;][]{2012SPIE.8446E..0JM},
both mounted on the 10-m Keck-I telescope (on Mauna Kea, Hawaii), in
$g$ and $R_s$ with LRIS and $J$ and $K_s$ with MOSFIRE.  All images 
were reduced in a standard fashion; for the P48 images this was done
using the IPAC pipeline \citep{2014arXiv1404.1953L}.  The P48 observed
the \sn\ field multiple times per night; to increase the depth of the
P48 data, we combined all images with seeing better than
3\arcsec\ over 3-day intervals before extracting the magnitude.  The
log of the imaging observations, where we list the 3-day averages for
the P48 data, is presented in Table~A.\ref{tab:logphotometry} in the
Appendix.

Aperture photometry of \sn\ was performed relative to two dozen
reference stars in the field with SDSS
\citep{1996AJ....111.1748F,2012PASP..124...62O} and 2MASS
\citep{2006AJ....131.1163S} photometry. We constructed the spectral
energy distribution of the reference stars in the field from SDSS
or 2MASS photometry, and derived their magnitudes in the relevant
filter by using the filter transmission curve; this significantly
reduces the scatter in the relative photometry in case the filter is
very different from that of the SDSS or 2MASS filters, such as the P48
$R$ band.  No attempt was made to subtract a reference image before
measuring the SN magnitude as the host turned out to be very faint:
$g_{\rm AB} \approx 27.0$ and $R_{s, \rm AB} \geq 26.0$ mag
(Sect.~\ref{sec:host}). The AB magnitudes listed in
Table~A.\ref{tab:logphotometry} are in the natural system of the
different filters, which are very similar to the SDSS/2MASS filter set
for all except the P48 $R$, P60 $B$, and Keck/LRIS $R_s$. The
magnitude uncertainty is a combination of the measurement error and the
uncertainty in the calibration derived from the scatter in the offsets 
with respect to the different reference stars.

Starting on 2013 April 8, long-slit low-resolution spectra of
\sn\ were obtained at several different epochs with the DEep Imaging
Multi-Object Spectrograph \citep[DEIMOS;][]{2003SPIE.4841.1657F}
mounted on the 10-m Keck-II telescope, with LRIS
\citep{1995PASP..107..375O} at Keck-I, and with the Dual Imaging
Spectrograph (DIS) mounted on the 3.5-m Astrophysical Research
Consortium (ARC) telescope at Apache Point Observatory (APO) in New
Mexico. These spectra, all taken with the slit aligned with the
parallactic angle \citep[see][]{1982PASP...94..715F}, were reduced,
extracted, telluric corrected, and flux calibrated in a standard
manner with IRAF\footnote{IRAF is distributed by the National Optical
  Astronomy Observatory, which is operated by the Association of
  Universities for Research in Astronomy (AURA), Inc., under
  cooperative agreement with the National Science Foundation (NSF).}
and custom IDL routines. Cosmic rays were rejected using the recipe
and routine of \citet{2001PASP..113.1420V}. The flux calibration was
performed using a standard star that was observed during the same
night in the same setting and slit width as the object, and with the
slit aligned with the parallactic angle.  If more than one standard
was observed during the night, we used the one with seeing conditions
nearest to those during the \sn\ observations, thereby minimizing the
slit-loss uncertainty. We note that an atmospheric dispersion
corrector was used when observing with LRIS, including the last epoch
spectrum which was taken at high airmass.

\begin{deluxetable}{lccrrccccc}
  \tabletypesize{\scriptsize}
  \rotate
  \tablecaption{Log of Spectroscopic Observations of \sn\label{tab:logspectroscopy}}
  \tablehead{
    \colhead{Date} & 
    \colhead{Telescope} & 
    \colhead{Instrument} & 
    \colhead{Exp.~Time} & 
    \colhead{Grating/Grism/Filter} & 
    \colhead{Slit~Width\tablenotemark{a}} &
    \colhead{$\lambda$ Coverage} & 
    \colhead{Resolution\tablenotemark{a,b}} & 
    \colhead{Seeing} &
    \colhead{Airmass} \\
    (UTC 2013) & & & (min.) & & \arcsec & (\AA) & \kms (\AA) & \arcsec & 
  }
  \startdata
  Apr.  8 & Keck\,2 & DEIMOS & 15          & 600ZD/GG455 & 0.7 & 4410--9640  & 150 (2.7)& 0.9 & 1.05\\ 
  Apr.  9 & Keck\,1 & LRIS   & 20          & 400/3400, 400/8500 & 0.7 & 3000--10,300 & 230 (4.3)& 0.9 & 1.05 \\ 
  Apr. 17 & VLT\,{\it Kueyen} & X-shooter & $4\times20$ & & 1.0/0.9/0.9 & 3000--24,800 & 55/34/59 & 0.8 & 2.13 \\ 
  May  8 & VLT\,{\it Kueyen} & X-shooter & $4\times20$ & & 1.0/0.9/0.9 & 3000--21,020\tablenotemark{c} & 55/34/59 & 0.7 & 2.13 \\ 
  May 10 & Keck\,1 & LRIS   & 10          & 600/4000, 400/8500 & 1.0 & 3130--10,260 & 190 (3.5)& 0.9 & 1.09 \\ 
  May 18\tablenotemark{d} & ARC\tablenotemark{d} &  DIS\tablenotemark{d} & 15 & B400/R300 & 1.5 & 3350--5450 & 250 (3.6) & 1.5 & 1.07 \\
  June  6 & Keck\,2 & DEIMOS & $2\times15$ & 600ZD/GG455 & 0.8 &  4410--9640 & 160 (3.0)& 0.9 & 1.05 \\ 
  June 10 & Keck\,2 & DEIMOS & $2\times10$ & 600ZD/GG455 & 1.0 &  4410--9630 & 180 (3.4)& 0.7 & 1.22 \\ 
  July 12 & Keck\,2 & DEIMOS & 20          & 600ZD/GG455 & 0.8 &  4900--10,140 & 150 (2.7)& 0.8 & 1.06 \\ 
  Sep.  9 & Keck\,1 & LRIS   & $2\times20$ & 400/3400, 400/8500 & 0.7 & 5600--10,230 & 240 (4.5) & 0.8 & 2.10 
  \enddata
  \tablenotetext{a}{The slit width and resolution (in units of
    \kms: $c \Delta\lambda/\lambda$) for the
    X-shooter observations is provided for the UVB/VIS/NIR arms.}
  \tablenotetext{b}{The resolution of the Keck DEIMOS and LRIS spectra
    has been determined at 5577~\AA; in parentheses it is listed in
    units of \AA.}
  \tablenotetext{c}{The May 8 X-shooter spectrum was taken with the
    K-band blocking slit, which improves the signal-to-noise ratio in
    the J and H bands, but limits the spectral coverage to below about
    21000~\AA.}
  \tablenotetext{d}{Owing to its low S/N, the
    ARC/DIS spectrum has not been used in the analysis.}
\end{deluxetable}

At two epochs close to the maximum $R$-band brightness ($R_{\rm
  AB}=20.28$ mag), \sn\ was also observed with the X-shooter echelle
spectrograph \citep{2011A&A...536A.105V} mounted on the {\it Kueyen}
8.2-m unit of the Very Large Telescope (VLT) of the European Southern
Observatory (ESO).  These observations were performed in Director's
Discretionary Time (DDT; ESO program ID: 291.D-5009) by staff
astronomers at Paranal in Chile. The echelle spectra of the object,
flux standards, and telluric standards were reduced to bias-subtracted,
flat-fielded, rectified, order-combined, and wavelength-calibrated 
two-dimensional spectra for all three arms (UVB, VIS, and NIR) using the 
Reflex package and version 2.2.0 of the X-shooter pipeline
\citep{2010SPIE.7737E..56M}. Subsequently, the object spectra were
optimally extracted, telluric corrected, and flux calibrated using
custom IDL routines.  The flux standard was observed with a slit much
wider slit (5\arcsec) than the object spectra (1\arcsec, 0.9\arcsec, and
0.9\arcsec\ for the UVB, VIS, and NIR arms, respectively). We
estimated the slit loss by fitting a one-dimensional Gaussian function 
to the object profile along the spatial direction every 300 pixels, or
6~\AA, along the dispersion axis. For each fit, we first sum 300
columns (assuming the spatial direction is along the image column) to
boost the object's signal to obtain a reliable Gaussian fit. The resulting 
best-fit values of the full width at half-maximum intensity (FWHM) as a
function of wavelength were approximated with a low-order polynomial,
and in comparison with the slit width the slit loss (again as a
function of wavelength) was estimated, for which the spectra were
corrected. The maximum slit losses in the UVB and VIS arms were 20\%
on April 17 and 13\% on May 8. 

Following the relative-flux calibration, all spectra were put on an
absolute flux scale by tying them to the $R$-band photometry; how this
was done is explained in more detail in Sect.~\ref{sec:spectra}.  The
wavelengths of all spectra were converted to vacuum and were
heliocentric corrected. The log of the spectroscopic observations is
shown in Table~\ref{tab:logspectroscopy}.

\section{Photometric evolution}
\label{sec:photometry}

\begin{figure}[h!]
  \centering
  \includegraphics[width=\hsize]{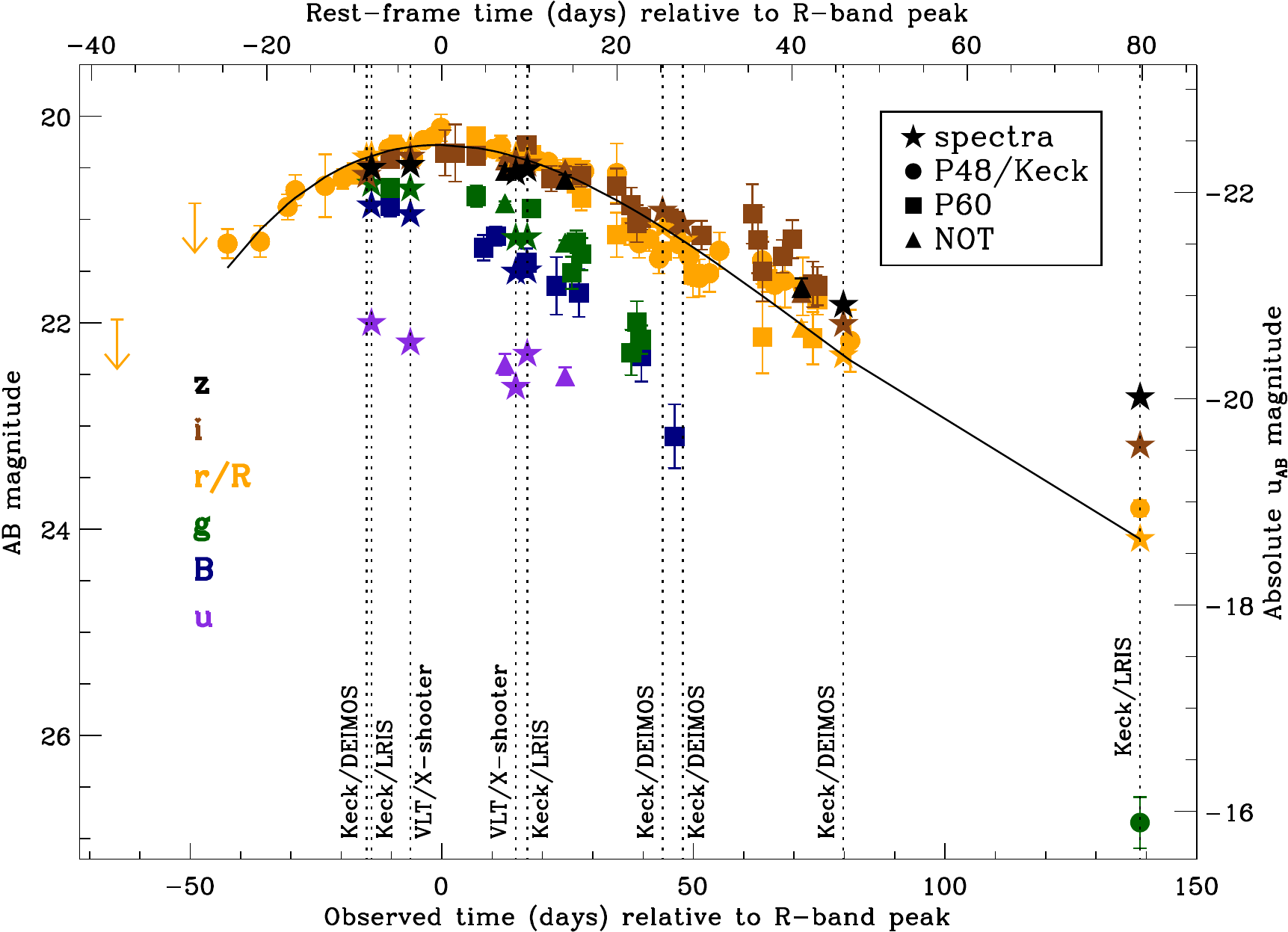}
  \caption{Light curves ($uBgRiz$) of \sn; the time in the observer's
    (rest) frame at the bottom (top) is relative to the maximum
    brightness in the $R$ band.  Filled circles, squares, and
    triangles are measurements from images obtained with the P48/Keck,
    P60, and NOT, respectively. For clarity, only a few prediscovery
    upper limits, determined using 3-day combined P48 data, are shown;
    see Table~\ref{tab:logphotometry} for all relevant magnitude
    limits.  The stars denote magnitudes derived from the
    flux-calibrated spectra after scaling them to the $R$-band light
    curve.  The times at which the spectra were obtained (see
    Table~\ref{tab:logspectroscopy}) are marked with the vertical
    dotted lines.  The right-hand axis shows the approximate absolute
    $u_{\rm AB}$ magnitude (corresponding to the $R$-band data), which
    includes a $K$ correction using the X-shooter spectrum taken on
    2013 April 17. A polynomial fit to the P48+Keck $R$-band data,
    depicted by the solid black line, indicates that a maximum
    absolute magnitude of $u_{\rm AB}=-22.5$ was reached on 2013 April
    23.6. \label{fig:lightcurves}}
\end{figure}

Figure~\ref{fig:lightcurves} shows the resulting light curves of
\sn. The AB magnitudes shown have been corrected for the Galactic
extinction along this sightline 
\citep[$E_{B-V}=0.012$~mag;][]{2011ApJ...737..103S}. The time axis shows the
number of days in the observer's frame (rest-frame days are given at
the top) relative to the peak time of the $R$-band light curve. The
latter has been determined to be 2013 April 23.6 (JD = 2456406.1)
from a low-order polynomial fit to the P48 and Keck $R$-band data;
this fit is shown by the solid black line in
Figure~\ref{fig:lightcurves}. The corresponding peak brightness is
$R_{\rm AB}=20.28$ mag. 

Following \citet{2002astro.ph.10394H}, we
determine the $K$ correction from observed $R_{\rm AB}$ to rest-frame
SDSS $u_{\rm AB}$ to be $K_{uR}=-0.57$ using the X-shooter spectrum
from 2013 April 17. We note that since the observed $R$ band
corresponds very well to rest-frame $u$ at $z=0.7403$, this $K_{uR}$
correction for \sn\ is very similar to $-2.5$\,log\,$(1+z)$. As the
\sn\ spectrum evolves with time, this $K$ correction changes from
$-0.55$ for the spectra taken at tens of days before peak, to $-0.8$
for the spectra taken a month or two after peak.  

Adopting a distances modulus at the \sn\ redshift ($z=0.7403$) of
$\mu=43.32$ mag and a $K$ correction of $K_{uR}=-0.57$, we derive an
absolute $u_{\rm AB}$-band peak magnitude of $M_{u,{\rm AB}}=-22.5$.
Transforming the observed $R_{\rm AB}$ magnitude to the rest-frame
Johnson $U$ and $B$ filters in the Vega system using the same spectrum
results in $M_{U,{\rm Vega}}=-23.5$ and $M_{B,{\rm Vega}}=-22.2$ mag.
These absolute magnitudes, which have an estimated uncertainty of
0.1--0.2~mag, are similar to those of the brightest Type I and II
SNSNe to date, such as SCP\,06F06 \citep[$M_{u,{\rm AB}}=-22.2$ mag,
  $z=1.189$;][]{2009ApJ...690.1358B}, PTF~09cnd \citep[$M_{u,{\rm
      AB}}=-22.1$ mag, $z=0.258$;][]{2011Natur.474..487Q}, SNLS 06D4eu
\citep[$M_{U,{\rm Vega}}=-22.7$ mag,
  $z=1.588$;][]{2013ApJ...779...98H}, and CSS121015:004244+132827
\citep[$M_{B,{\rm Vega}}=-22.6$ mag,
  $z=0.287$;][]{2013arXiv1310.1311B}. On the right-hand axis of
Fig.~\ref{fig:lightcurves} we indicate the approximate absolute
magnitude $M_{\rm u}$ corresponding to the $R$-band light curve of
\sn. Although this is correct for the values around peak, owing to the
evolving $K$ correction this is only an approximation at later times.

The date of explosion of \sn\ was estimated by fitting a parabolic and
exponential function to the pre-peak light curve, following
\citet{2014arXiv1404.4085O}. These two methods provide very similar
dates of explosion: March 2/3 (JD = 2456354/5), which is 51/52 days
(observed) before the $R$-band peak magnitude. In the rest frame, this
corresponds to an exponential rise time of about a month, which is
much longer than the rise times of standard core-collapse SNe of type
II \citep{2014ApJ...786...67A,2014arXiv1404.2004S} and type Ib/c
\citep{2014arXiv1408.4084T}. We caution that despite these two methods
leading to similar date estimates, the date of explosion is highly
uncertain. Moreover, the early-time (unobserved) light curve of
\sn\ may have contained a plateau as was observed for SLSN~2006oz
\citep{2012A&A...541A.129L}, in which case the rise time of about 30
rest-frame days should be considered a lower limit.

\section{Spectral evolution}
\label{sec:spectra}

\begin{figure}[h!]
  \centering
  \includegraphics[width=\hsize]{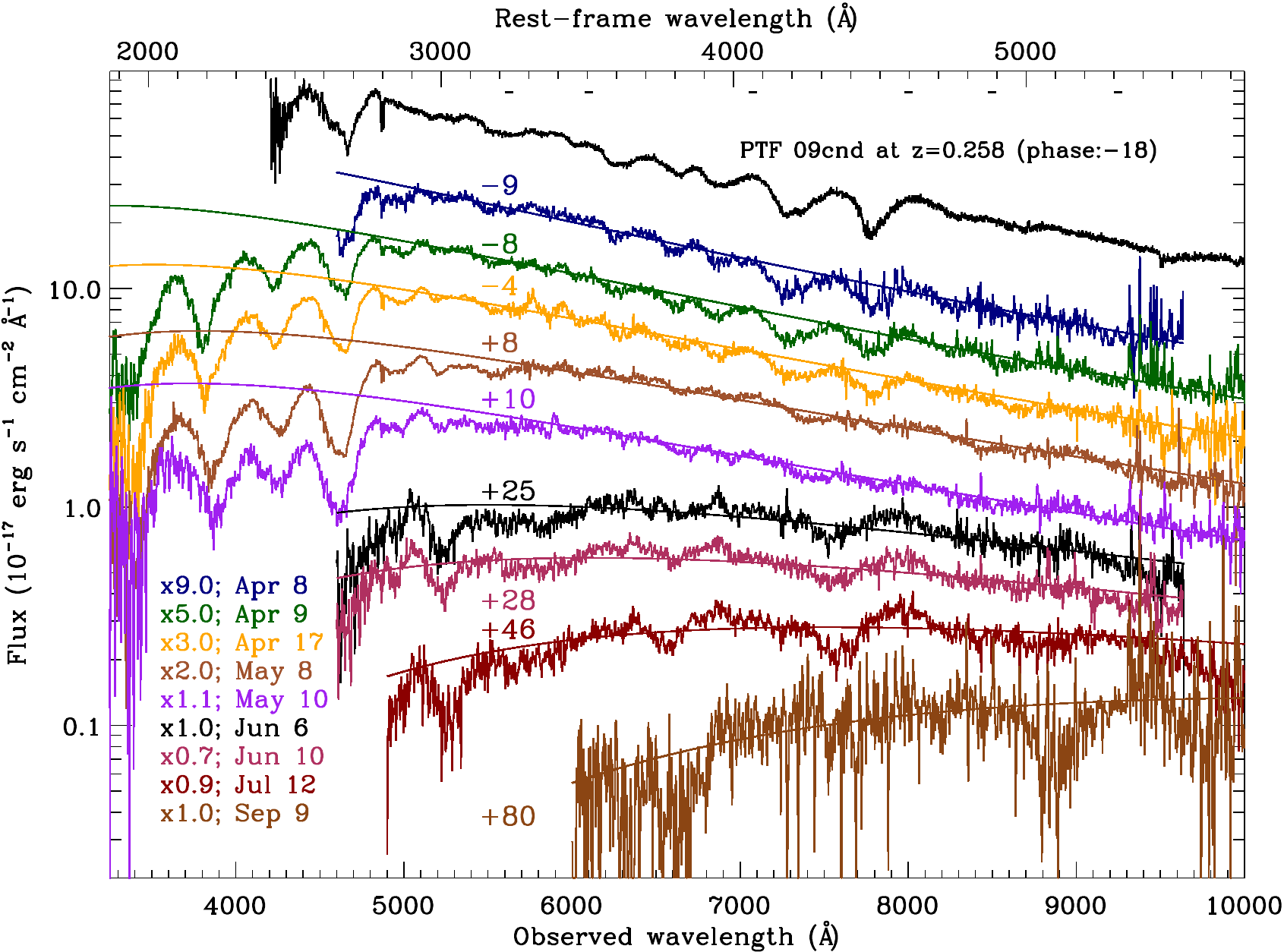}
  \caption{Time series of Keck/DEIMOS, Keck/LRIS, and VLT/X-shooter
    spectra of \sn\ (see Table~\ref{tab:logspectroscopy}).  The
    spectra were corrected for Galactic extinction
    \citep{2011ApJ...737..103S} and scaled to the polynomial fit to
    the $R$-band photometry depicted in Fig.~\ref{fig:lightcurves}. To
    avoid the spectra overlapping each other, an additional arbitrary
    scaling was applied as indicated at the bottom left, along with
    the civil date of observation (UTC 2013); the phase (rest-frame
    days relative to the $R$-band maximum) is shown next to each
    spectrum. The \sn\ spectra were smoothed with a Gaussian filter
    having a FWHM of 5~\AA. The spectra were fit with a Planck
    function to selected 50~\AA-wide wavelength regions (the same for
    all spectra; these regions are indicated with dashes at the top of
    the plot) free from apparent features, where the region blueward
    of rest-frame 3200~\AA\ has been discarded owing to the presence
    of very strong absorption. These blackbody fits are shown by the
    solid lines (see also Fig.~\ref{fig:bolometric}). For comparison,
    we include the Keck/LRIS spectrum of PTF~09cnd (shifted in
    wavelength to match that of \sn) at its actual flux scale
    \citep{2011Natur.474..487Q}.
    \label{fig:spectra}}
\end{figure}

We took particular care with the relative flux calibration of the
spectra (Sect.~\ref{sec:observations}) and placed them on an absolute
flux scale as follows. We multiplied the P48 $R$-band filter
transmission curve by each spectrum and integrated the resulting flux
over wavelength, did the same for the hypothetical AB standard star
(which emits 3631~Jy independent of wavelength), and scaled the
resulting AB magnitude to the value derived from the polynomial fit to
the $R$-band light curve at that epoch. Once on this absolute scale,
we derived synthetic $uBgriz$ magnitudes from each spectrum, provided
that the spectrum covered (most of) the wavelength range of the
corresponding transmission filter. These spectral magnitudes are shown
as stars in Figure~\ref{fig:lightcurves}. The agreement between these
synthetic magnitudes and the magnitudes derived from the images is
fairly good (with a scatter less than 0.2~mag), even for the X-shooter
spectra which were taken at high airmass, showing that the absolute
and relative flux calibration of the spectra is reasonably reliable.

Figure~\ref{fig:spectra} shows the time series of spectra of \sn,
starting at $9$ rest-frame days before the $R$-band peak and ending 
at a phase of $+80$ days. For clarity, some of
the spectra have been offset from their true absolute flux scale. At
the top, we also include the spectrum of PTF\,09cnd (shifted in
wavelength to match that of \sn), a prototypical hydrogen-poor
SLSN at $z=0.258$
\citep{2011Natur.474..487Q}.  The resemblance between \sn\ and
PTF\,09cnd (both lacking hydrogen features) combined with the peak
absolute magnitude leads to the classification of \sn\ as a SLSN-I.  
The redshift, as
measured from the narrow \ion{Mg}{1}, \ion{Mg}{2}, and \ion{Fe}{2}
absorption lines discussed in detail in Sect.~\ref{sec:narrowabslines}, 
is 0.7403. 

The spectra show very prominent broad absorption features in the
rest-frame near-UV range, as well as weaker
features in the rest-frame optical. The latter can be identified with
blueshifted \ion{O}{2} (the five features between $\lambda_{\rm
  rest}=3600$ and 4500~\AA) at early times \citep{2011Natur.474..487Q}.
At later times broad \ion{Ca}{2} ($\lambda_{\rm rest}\approx3700$~\AA)
and probably \ion{Fe}{2} ($\lambda_{\rm rest}\approx3700, 4300,
5000$~\AA) are present; that is, the spectrum of \sn\ is evolving to
appear similar to spectra of SNe~Ic 
\citep[see][]{2010ApJ...724L..16P,2011Natur.474..487Q}. The
identification of the near-UV features is not so unambiguous:
\citet{2011Natur.474..487Q} suggested that the lines at rest-frame
wavelengths of 2200~\AA, 2400~\AA, and 2650~\AA\ are produced by \ion{C}{2},
\ion{Si}{3}, and \ion{Mg}{2} (respectively), while
\citet{2013ApJ...779...98H} instead suggest them to be due to
\ion{C}{3}+\ion{C}{2}, \ion{C}{2}, and \ion{Mg}{2}+\ion{C}{2}. The
latter authors were also able to detect a feature at 1900~\AA\ in the
spectrum of SNLS~06D4eu, which they attribute to \ion{Fe}{3}, and it
was also suggested to be present in SLSN~2006oz
\citep{2012A&A...541A.129L}.

\begin{figure}
  \centering
  \includegraphics[width=0.6\hsize]{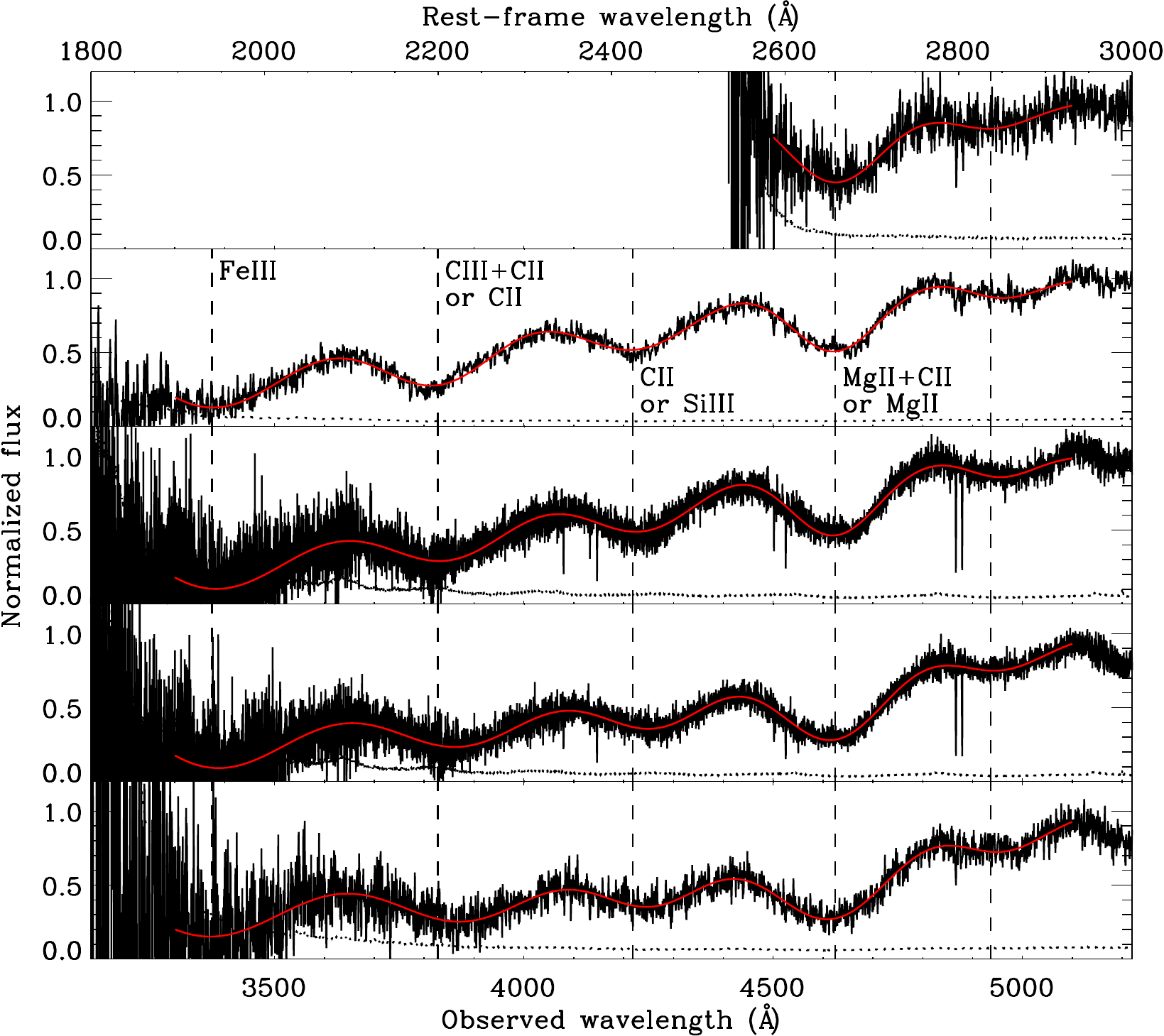}
  \caption{The blue side of the spectra taken at epochs 1--5 or phase
    $-9, -8, -4, +8$, and $+10$~days (top to bottom panels), normalized
    by the blackbody fits (see Fig.~\ref{fig:spectra}), are shown by
    the black solid lines. The spectra have not been smoothed.  The
    corresponding error spectra are indicated with the dotted
    lines. Gaussian fits are shown by the solid red lines. To illustrate 
    the evolution of the central wavelength of the lines with time, their
    best-fit wavelengths in the earliest epoch are marked with the
    vertical dashed lines.  The strengths of these features evolve
    little to moderately over the course of 20 rest-frame days.  The
    epoch 3 and 4 X-shooter spectra clearly display the narrow
    absorption features of \ion{Mg}{2} and \ion{Fe}{2} between
    4000~\AA\ and 5000~\AA\ (observed); these are less significant at
    the other epochs owing to the lower spectral resolution. In the
    second panel from the top we list the line identification of the
    broad features according to \citet{2013ApJ...779...98H} (top
    label) and \citet{2011Natur.474..487Q} (bottom label).
    \label{fig:normspectra}}
\end{figure}

To view these features in more detail, we zoom in on the near-UV part
of the relevant spectra in Fig~\ref{fig:normspectra}. These spectra
have been normalized by an approximation of the continuum obtained
from fitting a Planck function to selected 50~\AA-wide wavelength
regions (the same for all spectra) free from apparent features. The
region blueward of 5600~\AA\ (rest-frame 3200~\AA) has been
discarded in the fit owing to the presence of very strong
absorption. These Planck fits are shown by the solid lines in
Figure~\ref{fig:spectra}.

To derive basic quantities such as line center and strength, we fit
simple Gaussian profiles to all the obvious lines present in the
near-UV range. Keeping in mind the fact that the extrapolation of the
continuum fits to this wavelength range is fairly uncertain, it is
still interesting to note that the strength of the absorption features
does not seem to vary much with time. The only exception is the
feature at rest-frame 2650~\AA, whose strength seems to increase at
later times.  The line center is independent of the continuum
extrapolation. For the features at 2200~\AA\ and 2400~\AA, it evolves
roughly 3000~\kms\ to the red over the course of 20 rest-frame
days. By contrast, the line center is constant for the absorption at
1900~\AA\ (uncertain due to the low signal-to-noise ratio [S/N] at 
these wavelengths), 2650~\AA, and 2850~\AA.

Recently, an expansion velocity of $\sim$ 16,500~\kms\ was estimated
from the velocity difference between the narrow \ion{Mg}{2} lines and
the broad absorption feature for the SLSN PS1-11ap at $z=0.524$
\citep{2014MNRAS.437..656M}. This is similar to those found for other
SLSNe: velocities range from
10,000--20,000~\kms\ \citep[e.g.,][]{2011Natur.474..487Q,2011ApJ...743..114C,2013ApJ...770..128I}. Assuming
the feature at rest-frame 2650~\AA\ is \ion{Mg}{2}~\l 2800, the
expansion velocity for \sn\ is 15,500~\kms. In Sect.~\ref{sec:Lbol},
we present an estimate of the blackbody radius evolution of \sn\ as a
function of time, which appears to be well described by a linear
increase in time until at least 50 rest-frame days after peak. The
expansion velocity corresponding to this radius evolution is
11,500~\kms.

\section{Bolometric luminosity evolution}
\label{sec:Lbol}

\begin{figure}
  \centering
  \includegraphics[width=0.6\hsize]{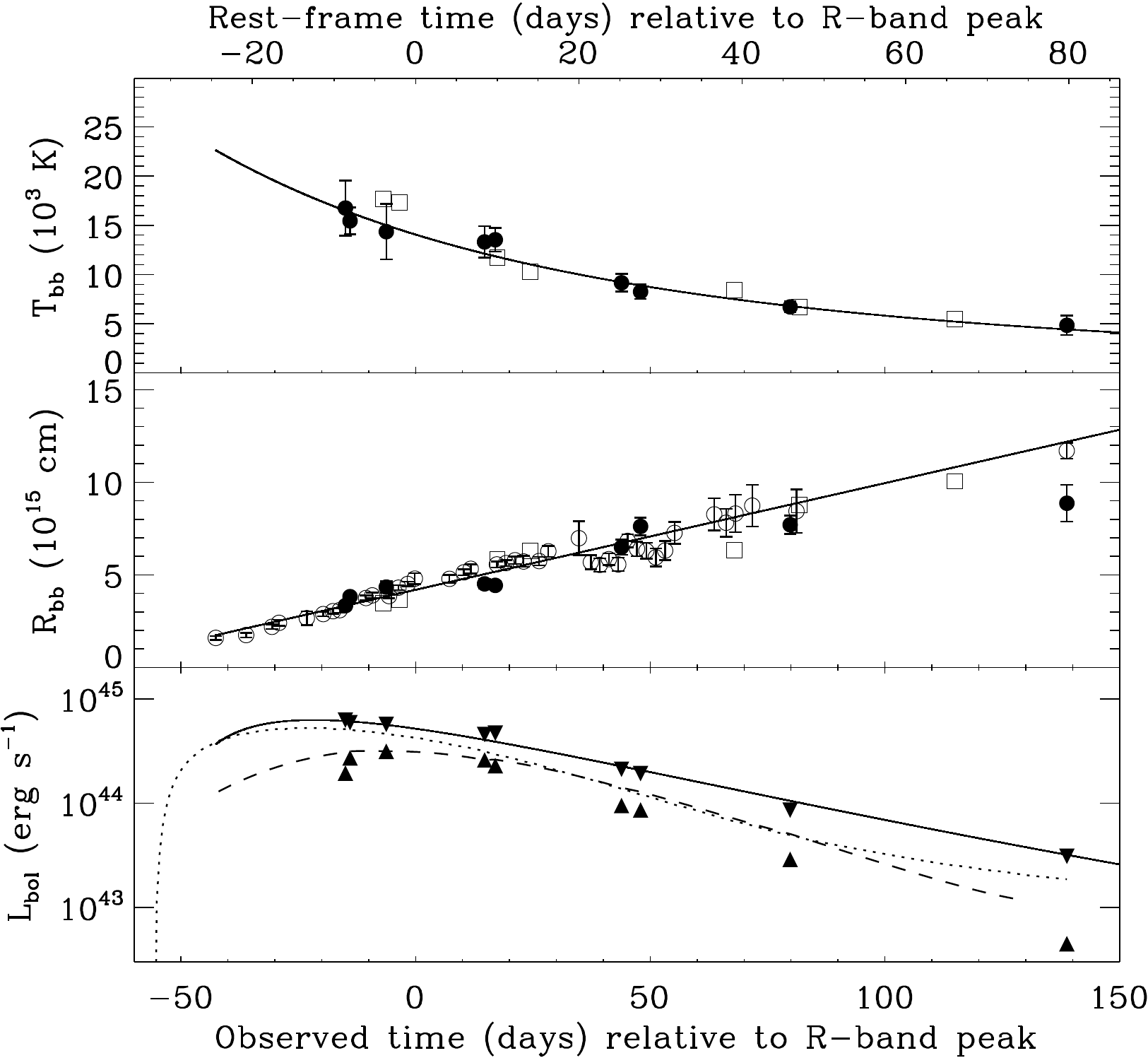}
  \caption{Time evolution of the blackbody temperature (top panel),
    blackbody radius (middle), and bolometric luminosity (bottom
    panel) of \sn. The blackbody fits to the observed spectra (shown
    by the solid lines in Fig.~\ref{fig:spectra}) provide nine
    temperature and radius estimates, depicted with the filled
    circles. The down-pointing triangles in the bottom panel indicate
    the bolometric luminosity ($L_{\rm bol}=4 \pi R^2 \sigma T^4$)
    corresponding to these estimates. The upward-pointing triangles
    show the luminosities inferred from simply integrating the
    observed spectra over wavelength and converting the integrated
    flux to luminosity. The functions fit to the temperature (power
    law) and radius (first-order polynomial) evolution combine to
    produce an estimate of the bolometric luminosity evolution of \sn,
    shown by the solid line in the bottom panel. The open circles in
    the middle panel are the blackbody radii inferred from the
    P48/Keck $R$-band photometry and assuming the power-law
    temperature evolution; these points are included in the polynomial
    fit to the radius evolution. The open squares indicate the $T$ and $R$
    evolution reported for the SLSN CSS121015
    \citep{2013arXiv1310.1311B}; these points are not used in the
    fits. The dashed line in the bottom panel, our best estimate for
    the bolometric light curve, is a combination of interpolation
    between the available spectra, and extrapolation using the
    blackbody fits, which are modified in the near-UV by the Gaussian
    fits to the broad absorption features (see
    Fig.~\ref{fig:normspectra}). Finally, the dotted line in the
    bottom panel indicates the light curve produced by a magnetar
    model with a magnetic field of $B=2.3\times10^{14}$~G and an
    initial spin period of $P=1.1$~ms. \label{fig:bolometric}}
\end{figure}

As already mentioned in the previous section, the solid lines in
Figure~\ref{fig:spectra} depict blackbody fits to selected 50~\AA-wide
wavelength spectral regions (the same for all spectra) redward of
5600~\AA\ (to avoid the strong absorption features in the blue) and
free from apparent features. These blackbody fits result in an
estimate of the temperature and radius of an expanding photosphere at
each of the nine spectral epochs; these are shown by the solid circles
in the top (temperature) and middle (radius) panels of
Figure~\ref{fig:bolometric}. For comparison, the open squares show the
temperatures and radii as derived for the SLSN CSS121015:004244+132827
at $z=0.287$ by \citet{2013arXiv1310.1311B}; these data points are not
used in any of the fits described below.

The temperature evolution of \sn\ is poorly described by a linear
decline in time; instead a second-order polynomial and a power-law
function ($T\propto(t-t_0)^{\alpha}$) provide a good fit. We adopt the
power-law function, as the polynomial fit results in an unphysical
upturn at late times which is avoided in the power-law fit. The
best-fit values for the power-law fit parameters are highly dependent
on the choice of the start time $t_0$. For example, when fixing $t_0$
to $-30$ rest-frame days, $\alpha=-0.1$, with a very steep decay at
early times, whereas $\alpha=-3.5$ when $t_0$ is unconstrained.  The
latter fit is shown by the solid line in the top panel of
Figure~\ref{fig:bolometric}. The photospheric radius evolution is well
described by a linear increase in time, with a slope of
$1.0\times10^{14}$~cm per rest-frame day, or 11,500~\kms. Adopting the
temperature evolution inferred above, the radius evolution can also be
constrained from the imaging measurements by converting the magnitudes
to flux at the effective wavelength of the corresponding filter. In
the fit to the radius evolution, we have also included the P48/Keck
$R$-band light-curve data, which are shown by the open circles in the
middle panel of Fig.~\ref{fig:bolometric}. We note that at phase +80,
it is not clear whether the photometry or spectroscopy provides a more
reliable estimate of the radius; it may well be that the radius
evolution is flattening off around this phase.

These fits to the photospheric temperature and radius evolution are
combined to estimate the total bolometric luminosity of \sn, using
$L_{\rm bol}=4 \pi R^2 \sigma T^4$, illustrated by the solid line in 
the bottom panel of Figure~\ref{fig:bolometric}. The downward-pointing
triangles in this panel show the bolometric luminosity corresponding
to the temperatures and radii inferred from the nine spectra.  A
strict lower limit can be obtained by simply integrating the observed
spectra over their observed wavelength ranges, and converting the
integrated flux to luminosity in the rest frame. Those values are
indicated by the upward-pointing triangles in the bottom panel.

Finally, we provide a ``best'' estimate of the bolometric light curve
of \sn\ by constructing a flux surface as a function of time and
wavelength as follows. Whenever the flux at a particular wavelength at
a particular time can be recovered by interpolation between the
observed spectra, we do so. Before interpolation, the spectra are
first median filtered to avoid introducing extreme features based on a
single low or high value in one of the spectra. Each interpolated
spectrum is scaled to the value of the polynomial fit to the P48/Keck
$R$-band photometry (shown by the solid line in
Figure~\ref{fig:lightcurves}). For wavelengths and time intervals
falling outside the surface covered by the observed spectra, we adopt
the blackbody model flux described by the temperature evolution power
law, again combined with the P48/Keck $R$-band flux evolution.  In the
UV part, this flux surface is modified by the near-UV absorption
features as described by the Gaussian fits discussed in the previous
section. We simply extrapolate these fits to earlier and later times,
assuming their strength at early times is the same as in the spectrum
at epoch 2, and the same as in the epoch 5 spectrum for late times.
This flux surface ranges in time from $-24$ to $+80$ days, and in
wavelength from 1900 to 12,000~\AA, both in the rest frame; beyond these
boundaries it is assumed to be zero. The integral of this flux surface
over wavelength results in the dashed line in the bottom panel of
Figure~\ref{fig:bolometric}. As the luminosity absorbed by the UV
features is expected to be re-emitted at longer wavelengths, our
``best'' bolometric luminosity estimate (which has been corrected
downward owing to the strong absorption in the UV) does not necessarily
describe the true bolometric luminosity evolution of \sn. We expect
the latter to be bracketed by the blackbody and the ``best''
luminosity curves, depicted by the solid and dashed lines in the
bottom panel of Figure~\ref{fig:bolometric}, respectively.

The observed bolometric peak luminosity of \sn\ is $3.2\times
10^{44}$~erg~\persec\ ($6.3\times 10^{44}$~erg~\persec) assuming the
``best'' (blackbody) bolometric light curve. Integrating these two
light curves over the time interval $-24$ to $+80$ rest-frame days
leads to the following estimates of the total radiated energy of \sn:
$1.3\times10^{51}$~erg (``best'') and $2.5\times10^{51}$~erg
(blackbody). These values are similar to those found for other
SLSNe, such as the Type I SNLS 06D4eu \citep{2013ApJ...779...98H} and
the Type II CSS121015:004244+132827 \citep{2013arXiv1310.1311B}.

Closely following \citet{2013ApJ...770..128I}, using their equations
D1 through D7, we generated light curves using a magnetar model based
on the \citet{1982ApJ...253..785A} formalism. We found that a
magnetar with $B=2.27\times10^{14}$~G, $P=1.14$~ms,
$\tau_m=34.59$~days, and $t_i=-32.06$~days is consistent with our
estimated range for the bolometric light curve of \sn, as shown by the
dotted line in the bottom panel of Figure~\ref{fig:bolometric}.

\section{Narrow absorption features}
\label{sec:narrowabslines}

\begin{figure}
  \centering
  \includegraphics[width=0.6\hsize]{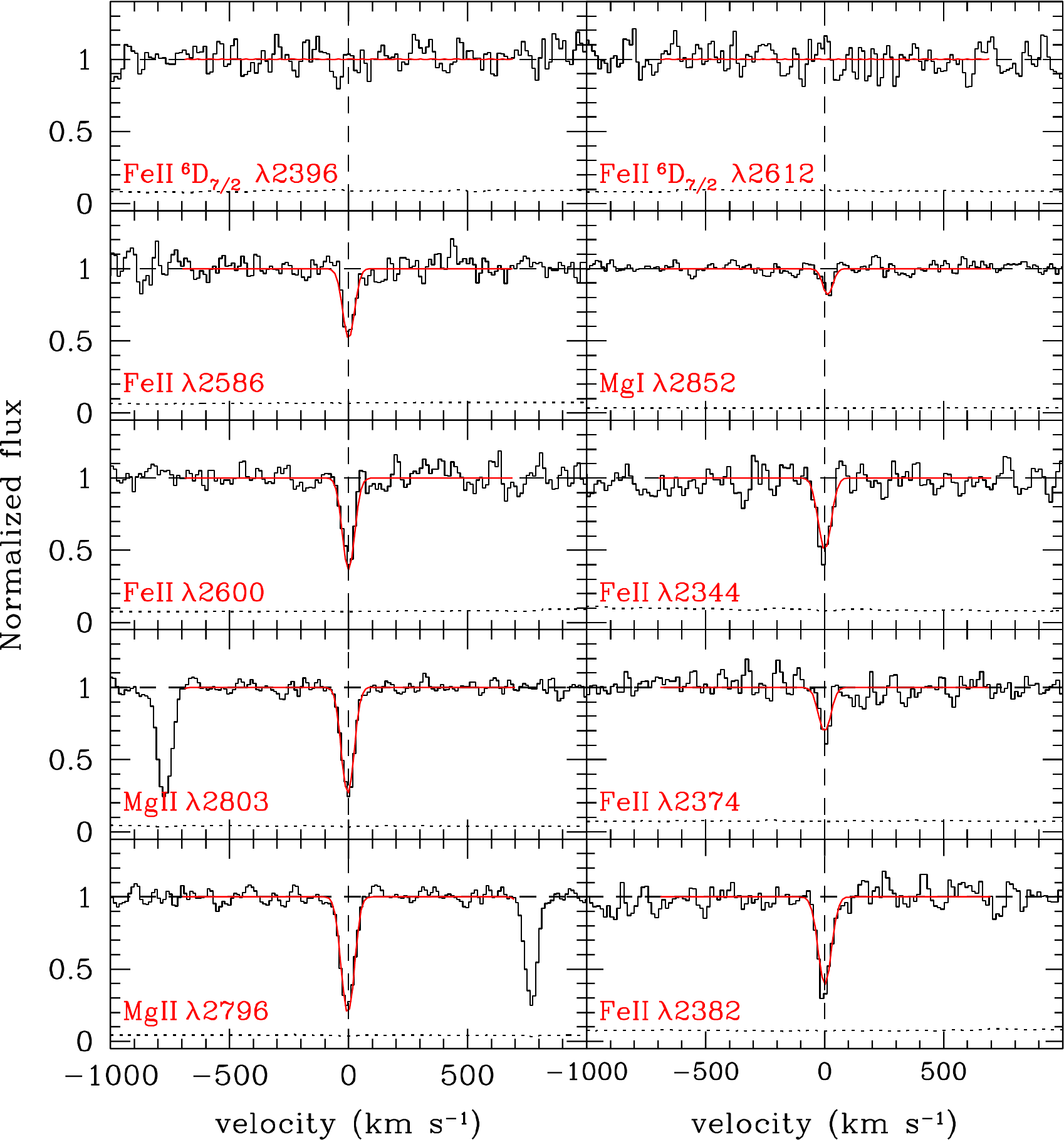}
  \caption{Spectral regions centered on absorption lines of
    \ion{Mg}{1}, \ion{Mg}{2}, and \ion{Fe}{2}, associated with neutral
    gas, in the averaged X-shooter spectrum (black solid line). No
    other narrow absorption lines are detected. The spectrum has been
    normalized to unity by fitting a low-order polynomial to the
    surrounding continuum.  The red solid line indicates the
    best-fitting Voigt profile, from which we infer the column
    densities (Table~\ref{tab:absorption}).  In the top panels, we
    also show the expected location of the strongest transitions of
    the first excited level of \ion{Fe}{2}, $^6$D$_{7/2}$, which are
    not detected. \label{fig:narrowlines}}
\end{figure}

The combination of X-shooter's sensitivity and intermediate resolving
power, and the brightness and significant redshift of \sn, resulted in the
clear detection of several narrow rest-frame near-UV absorption
lines. No significant variation of the absorption-line strength is
detected when comparing the spectra taken on April 17 and May
8. Spectral regions around the relevant absorption lines detected in
the combined spectrum are presented in Figure~\ref{fig:narrowlines}. The
continuum has been normalized to unity, and a redshift of 0.7403
has been adopted as the systematic redshift of the host galaxy of \sn.
In Table\,\ref{tab:absorption} we report the observed transitions,
wavelengths, and rest-frame equivalent widths ($W_r$), plus some limits
for other nondetected significant transitions. These include excited
lines associated with iron and nickel, regularly detected in GRB
afterglow spectra \citep[e.g.,][]{Chen05,2007A&A...468...83V}.

\begin{deluxetable}{lccccccc}
  \tablecaption{Absorption-Line Parameters\label{tab:absorption}}
  \tablehead{
    \colhead{Ion} & 
    \colhead{$\lambda_{\rm obs}$} & 
    \colhead{$W_{\rm r}$} & 
    \multicolumn{2}{c}{COG} &&
    \multicolumn{2}{c}{Voigt} \\
    &
    \colhead{(\AA)} & 
    \colhead{(\AA)} & 
    \colhead{$\log N_{\rm X}$ [\cm]} & 
    \colhead{$b$ (\kms)} &&
    \colhead{$\log N_{\rm X}$ [\cm]} & 
    \colhead{$b$ (\kms)}
  }
  \startdata
      Fe\,\textsc{ii} $\lambda2600$ & 4525.08 & $0.293\pm0.029$ & $14.04^{+0.13}_{-0.07}$ & $12.64\pm0.91$ && $14.25\pm0.10$ & $11.9\pm1.1$   \\ 
      Fe\,\textsc{ii} $\lambda2586$ & 4501.54 & $0.226\pm0.025$ &  . . . & . . . && . . . & . . .  \\
      Fe\,\textsc{ii} $\lambda2382$ & 4146.72 & $0.354\pm0.027$ &  . . . & . . . && . . . & . . . \\
      Fe\,\textsc{ii} $\lambda2374$ & 4132.27 & $0.134\pm0.027$ & . . . & . . . && . . . & . . .  \\
      Fe\,\textsc{ii} $\lambda2344$ & 4079.63 & $0.261\pm0.030$ & . . . & . . . && . . . & . . .  \\
      \tableline
      Mg\,\textsc{ii} $\lambda2796$ & 4866.44 & $0.510\pm0.015$ & $14.44^{+0.25}_{-0.20}$ & (12.64)\tablenotemark{a} && $14.68^{+0.34}_{-0.28}$ & (11.9)\tablenotemark{a}  \\
      Mg\,\textsc{ii} $\lambda2803$ & 4878.93 & $0.451\pm0.014$ & . . . & . . . && . . . & . . .  \\
      Mg\,\textsc{i} $\lambda2852$ & 4965.26 & $0.076\pm0.014$ & $11.82\pm0.09$ &  (12.64)\tablenotemark{a} && $11.94\pm0.06$ & (11.9)\tablenotemark{a} \\ 
      \tableline
      Mn\,\textsc{ii} $\lambda2594$ & 4516.94 & $<0.082$ &  . . . & . . . && $<12.80$ & (11.9)\tablenotemark{a}  \\
      Mn\,\textsc{ii} $\lambda2576$ & 4484.53 & $<0.070$ &  $<12.2$ & (12.64)\tablenotemark{a} && . . .  & . . .  \\
      Mn\,\textsc{ii} $\lambda2606$ & 4536.02 & $<0.092$ & . . . & . . . && . . .  & . . . \\
      \tableline
      Fe\,\textsc{ii}$^* \lambda2612$ & 4546.80 & $<0.101$ & . . . &  . . . && . . . &  . . .   \\
      Fe\,\textsc{ii}$^* \lambda2396$ & 4170.38 & $<0.095$ & $<12.4$ & (12.64)\tablenotemark{a} &&  . . .  &  . . .   \\
      Fe\,\textsc{ii}$^* \lambda2626$ & 4570.81 & $<0.090$ & . . . &  . . . && $<12.8$ & (11.9)\tablenotemark{a}  \\
      Fe\,\textsc{ii}$^* \lambda2389$ & 4158.20 & $<0.083$ & . . . &  . . . &&  . . .  &  . . .   \\
      Fe\,\textsc{ii}$^* \lambda2365$ & 4116.77 & $<0.077$ & . . . &  . . . &&  . . .  &  . . .   \\
      Fe\,\textsc{ii}$^* \lambda2333$ & 4061.02 & $<0.111$ & . . . &  . . . &&  . . .  &  . . .   \\
      \tableline
      Fe\,\textsc{ii}$^{5*} \lambda2360$ & 4108.36 & $<0.078$ & $<13.54$ & (12.64)\tablenotemark{a} && $<14.0$ & (11.9)\tablenotemark{a} \\
      Fe\,\textsc{ii}$^{5*} \lambda2348$ & 4087.68 & $<0.094$ & $<13.42$ & (12.64)\tablenotemark{a} && $<13.9$ &  (11.9)\tablenotemark{a} \\
      Fe\,\textsc{ii}$^{5*} \lambda2332$ & 4058.42 & $<0.107$ & . . . & . . . && . . . & . . . \\
      \tableline
      Ni\,\textsc{ii}$^{**} \lambda2316$ & 4031.84 & $<0.120$ &  $<13.06$ & (12.64)\tablenotemark{a} && $<13.4$ & (11.9)\tablenotemark{a} \\
      Ni\,\textsc{ii}$^{**} \lambda2217$ & 3858.54 & $<0.234$ &  $<13.18$ & (12.64)\tablenotemark{a} && . . . & . . .
      \enddata
      \tablenotetext{a}{Assumed Doppler width (including uncertainty
        of $\sim 1$~\kms) to estimate column density or upper limit.}
\end{deluxetable}

We performed Voigt profile fitting of the absorption lines to infer
the column densities. For completeness, we explored the lines'
parameters by using two other different methods: the apparent optical
depth \citep[AOD;][]{1991ApJ...379..245S} and the curve of growth
\citep[COG;][]{1978ppim.book.....S}. Although we are not in the ideal
situation of high spectral resolution ($R \gsim 40,000$) and
signal-to-noise ratio ($S/N>30$), comparing results from these
different tools allows us to evaluate the reliability of the resulting
column densities. These are summarized in Table\,\ref{tab:absorption}
for the COG and Voigt profiles; the AOD results do not provide
additional information so they are not included in the table. Adopting
the Voigt profile values, we find log $N$(\ion{Mg}{1}) $=11.94\pm0.06$,
log $N$(\ion{Mg}{2}) $=14.7\pm0.3$, and log
$N$(\ion{Fe}{2}) $=14.25\pm0.10$.  The relative magnesium over iron
abundance is [Mg/Fe] $=0.3\pm0.3$.  In principle, this quantity can be
used to infer the amount of dust in a similar way as is done for
[Zn/Fe] \citep[see][]{2013A&A...560A..88D}; however, the large uncertainty
in the observed [Mg/Fe] prevents us from drawing any conclusions about the
dust depletion.

The absorption lines are in one single and narrow component. Being
narrow is an indication that the absorbing gas is not associated with
the supernova ejecta, for which features are generally very broad and
blueshifted with respect to the rest-frame velocity (see,
e.g., Fig.~\ref{fig:spectra}). In Sect.~\ref{sec:excitation} we derive
a lower limit on the distance between the \sn\ and the absorbing gas
responsible for the narrow lines, showing that these lines are produced 
by gas in the interstellar medium (ISM) of the host galaxy of \sn.

We determined a small {\it effective} Doppler width, $b\approx12$~\kms,
in a single component, indicating that the low-ionization gas is
distributed in a region with a small velocity dispersion along the
sight line.  The low ionization state is indicated by the presence of
\ion{Mg}{1}, \ion{Mg}{2}, and \ion{Fe}{2}. Owing to the limited
resolution and resolving power, we cannot exclude that the intrinsic
absorption is originating in more than one cloud.

However, we still can limit these clouds to be distributed over a
velocity range smaller than $\sim 100$~\kms.  Following
\citet{Ledoux06} \citep[see also][]{2013MNRAS.430.2680M}, we measure
the line profile velocity width, $\Delta V$, defined by the
wavelengths where the cumulative apparent optical depth profile is
equal to 5\% and 95\%: $\Delta V = c
[\lambda(95\%)-\lambda(5\%)]/\lambda_0$~\kms, where $c$ is the speed
of light. We use the absorption lines \ion{Fe}{2}~\l\l2374,2586,2600
and \ion{Mg}{2}~\l2803 and find $\Delta V \approx 76$~\kms. This value
is only moderately higher than the resolving power in the UVB arm of
X-shooter (55~\kms), which implies that the true $\Delta V$ of
\sn\ could be lower. Moreover, the absorption lines used may be
somewhat saturated, which would also lead to an overestimate of the
intrinsic $\Delta V$ value. Compared to the QSO damped \lya\ absorber
(DLA) sample of \citet{Ledoux06}, the \sn\ absorber is among the 30\%
lower velocity DLAs, even though the UVES resolution of the Ledoux et
al. sample allows the $\Delta V$ to be constrained down to much lower
velocities (20~\kms\ or so).  Using the QSO-DLA velocity-metallicity
relations as shown in the left panel of Figure~4 of \citet{Ledoux06},
and considering the \sn\ $\Delta V$ that we measure as an upper limit,
we can obtain a very crude limit on the metallicity along the
\sn\ sightline: [M/H] $\lesssim -1$. One additional caveat in this
comparison is that the low-redshift QSO-DLA sample of \citet{Ledoux06}
is at $1.7<z<2.43$, while the redshift of \sn\ is much lower,
$z=0.7403$. Using the redshift evolution of the DLA mass-metallicity
relation as measured by \citet{2013MNRAS.430.2680M} (see their Figure
2), we estimate [M/H] $\lesssim -0.7$ for the host galaxy of \sn.

\section{Distance between \sn\ and \ion{Fe}{2} gas}
\label{sec:excitation}

We estimate a lower limit on the distance between the supernova and
the \ion{Fe}{2} and \ion{Mg}{2} material from the absence of
fine-structure lines, similar to what has been done for GRB afterglows
\citep[e.g.,][]{2007A&A...468...83V}.  The ultraviolet photons from
the GRB afterglow or SLSN will excite any \ion{Fe}{2} material
that is near enough. At sufficient photon fluxes, a significant
fraction of the atoms will be excited to levels just above the ground
state. In the case of GRB afterglows, absorption lines from these
excited levels of ions such as \ion{Si}{2}
\citep{vrees030323,2012MNRAS.420..627S}, \ion{Fe}{2}
\citep{Chen05,2006astro.ph..1057P}, and \ion{Ni}{2} have been detected
along tens of sightlines. The GRB-cloud distances inferred from these
observations range from about 50~pc to well over a kiloparsec
\citep[e.g.,][]{2007A&A...468...83V,D'Elia09a,080310paper2}.

To our knowledge, the only SN in which such fine-structure
lines have been detected to date is the Type IIn SN~1998S
\citep{2000ApJ...536..225B}. Their detection is challenging in the
case of SNe, as spectroscopic observations at intermediate
resolving power or higher in the near-UV are required. SLSNe, being
very luminous and UV bright, in principle allow the detection of these
fine-structure lines from the ground, as they are sufficiently bright
to obtain a good S/N at the relevant optical wavelengths.

\begin{figure}
  \centering \includegraphics[width=0.6\hsize]{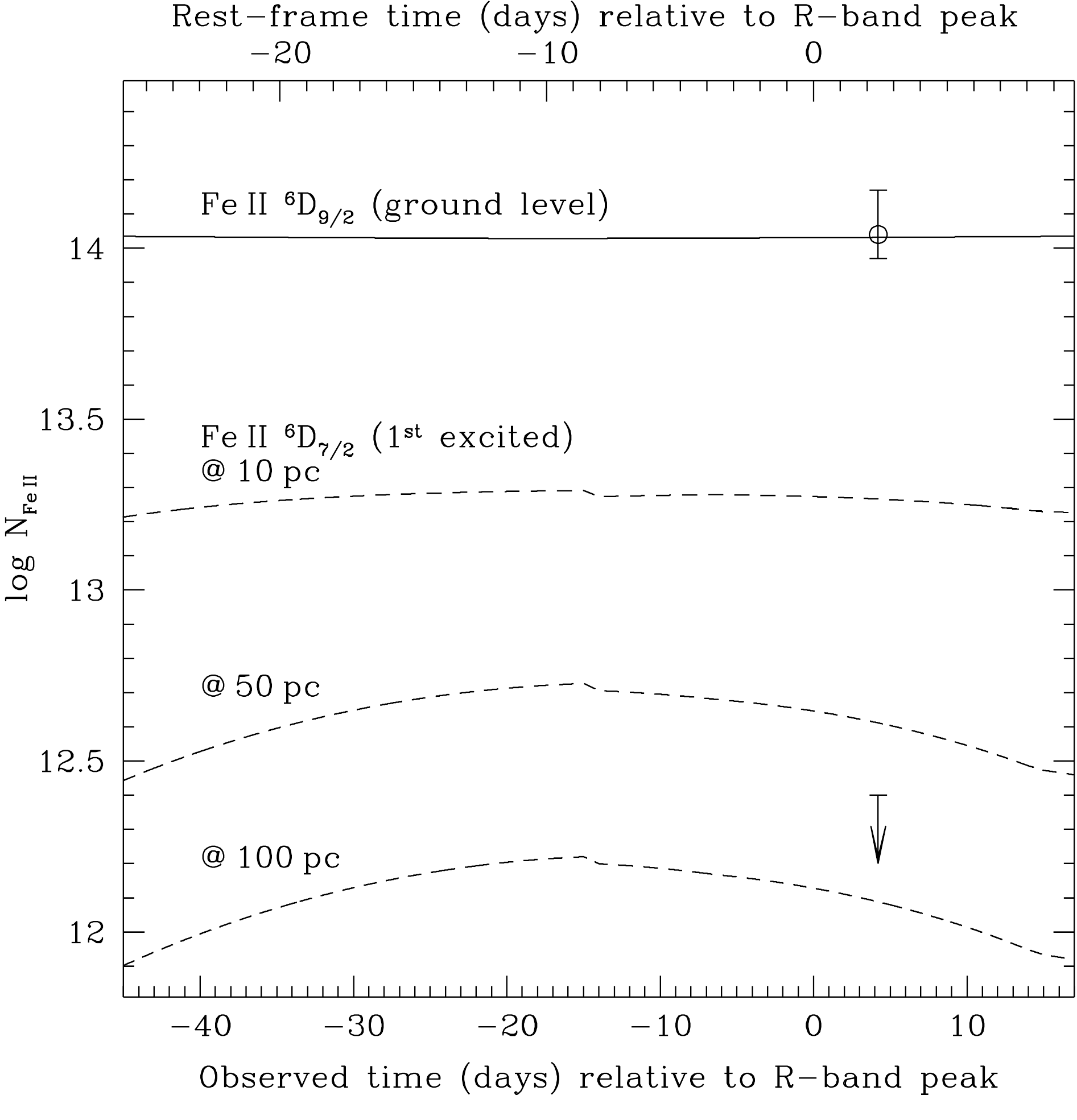}
  \caption{Expected population of the first excited level of
    \ion{Fe}{2} ($^6$D$_{7/2}$, indicated with the dashed lines) due
    to excitation by UV photons released by \sn, if the \ion{Fe}{2}
    atoms were located at a distance of 10~pc, 50~pc, and 100~pc from
    the UV source. The upper limit on the column density, log
    $N$(\ion{Fe}{2} $^6$D$_{7/2}$) $<$ 12.4, shows that the \ion{Fe}{2}
    atoms (as detected by the narrow absorption lines) are at least
    50~pc away from \sn. \label{fig:excitation}}
\end{figure}

For \sn, we measured a column density for \ion{Fe}{2} of log
$N$(\ion{Fe}{2}) $=14.25\pm0.10$, while constraining the population in
the first excited level to be log $N$(\ion{Fe}{2} $^6$D$_{7/2}$) $<$
12.4.  We calculate the excitation of the \ion{Fe}{2} ions, closely
following the method described by \citet{080310paper2}, assuming
different distances between \sn\ and the absorbing material.  This
provides the population of different excited levels, including the
first excited level $^6$D$_{7/2}$, as a function of time.  For the
input light curve, we adopt the ``best'' flux evolution as a function
of time and wavelength of \sn\ that was described in
Sect.~\ref{sec:Lbol}, starting at 30 rest-frame days before the
$R$-band peak. We note that this does not include a potential
early-time UV flash; if present, such a flash would increase the
amount of excitation and would therefore increase the lower limit on
the distance inferred below. Since the amount of ionizing flux at
rest-frame wavelengths below 912~\AA\ from \sn\ is not expected to be
high due the large opacity of the ejected material in the UV
wavelength range, even at early times (see Fig.~\ref{fig:spectra}), we
neglect ionization in the modelling. Even in case \sn\ would be
emitting a considerable amount of ionizing photons, the estimated
\ion{H}{1} column density associated with the metal absorption lines
is of the order of log N(\ion{H}{1}) $\approx$ 20 (see
Sect.~\ref{sec:discussion}), which is sufficient to shield the
absorber from most photons capable of ionizing \ion{Fe}{2}.

The \ion{Fe}{2} excitation is caused by UV photons with wavelengths from
the Lyman limit up to about 2600~\AA. We conservatively assume that
the \sn\ flux is negligible for rest-frame wavelengths below
1900~\AA. If there is significant flux present in that wavelength
region, the inferred lower limit on the distance would increase, since
the amount of excitation would be underestimated. The dashed lines in
Figure~\ref{fig:excitation} show the expected time evolution of the
\ion{Fe}{2} $^6$D$_{7/2}$ level, assuming a distance of 10~pc (top
dashed line), 50~pc (middle), and 100~pc (bottom). For all three
calculations, the ground-level population is indicated with the same
solid line. The upper limit on the $^6$D$_{7/2}$ level column density
constrains the distance between \sn\ and the \ion{Fe}{2} absorption
that we observe to be at least 50~pc. If the \ion{Fe}{2}
ions were much closer, we should have detected significant
absorption lines from this level.

\section{The host galaxy of \sn}
\label{sec:host}

\begin{figure}
  \centering
  \includegraphics[width=0.6\hsize]{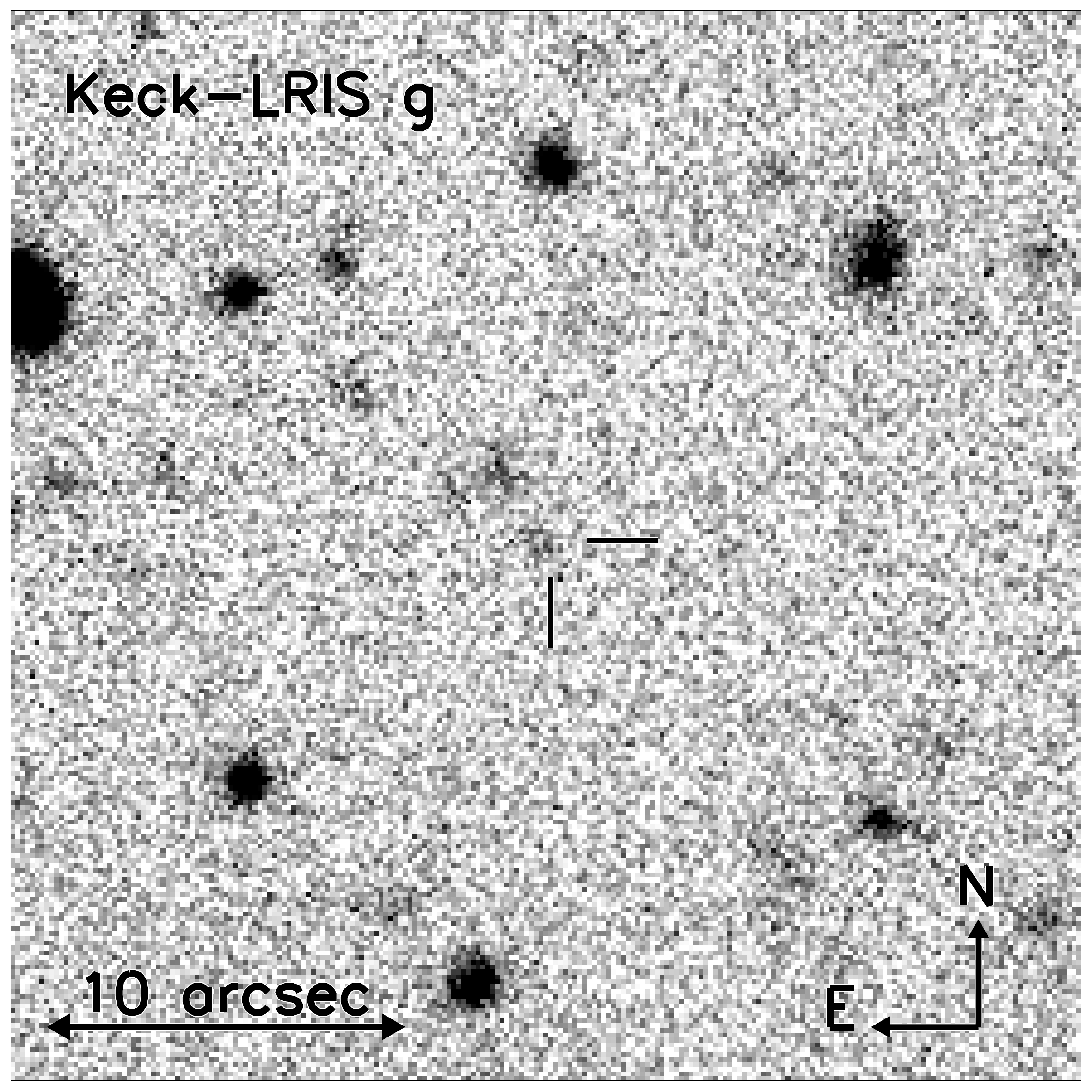}
  \caption{Keck/LRIS $g$-band image of the host galaxy of \sn; it
    is the sum of the 2013 September 9 and 2014 July 30
    observations. The position of the early-time supernova, indicated
    at the center, has been projected onto this image using NOT images
    from 2013 May 5 and several stars in the field for the
    alignment; the uncertainty in the projection is 0.4 pixels
    (0.05\arcsec), too small to show.  We measure the host-galaxy
    magnitude to be $g_{\rm AB} \approx 27.0$.
    \label{fig:hostimages}}
\end{figure}

We secured late-time images of the field of \sn\ with Keck/LRIS on
2013 September 9, 2014 April 28, and July 30 using filters $g$
and $R_s$ on all three nights.  These dates correspond to 80, 213, and
267 rest-frame days after the $R$-band peak magnitude. In $g$,
the magnitudes at these epochs are $26.89\pm0.25$, $26.49\pm0.14$,
and $>26.9$, respectively (all AB; see
Table~\ref{tab:logphotometry}). The bright April measurement may have
been due to a rebrightening of the supernova. Although the host is
formally not detected in the July measurement, it appears to be
present at the 2.5$\sigma$ level. Summing the September 9 and July 30
images, we find $g=26.8\pm0.2$ mag; this combined image is shown in
Figure~\ref{fig:hostimages}. From the Keck $R_s$-band images, we measure
$23.83\pm0.08$, $24.84\pm0.09$, and $26.01\pm0.22$ mag at 80, 213,
and 267 days, respectively (all AB; see
Table~\ref{tab:logphotometry}). Even in the last $R$-band epoch the SN
may still be contributing to the flux. The $g$-band detection ($g_{\rm AB} 
\approx 27.0$ mag) and the limiting $R_s$-band measurement ($R_{s,\rm AB}
\geq 26.0$ mag) roughly correspond to an absolute $B$-band magnitude
$M_{B,{\rm Vega}} \gtrsim -17.7$ for the host of \sn. On 2014
February 13, or a phase of 170 days, we also imaged the field with
DCT/LMI in the SDSS $r$ band. We do not detect \sn\ or its host, and
derive a limiting magnitude of $r>25.12$. In addition, we imaged the
field of \sn\ in the near-infrared with Keck/MOSFIRE on 2014 June
7 in $K_s$ and on 2014 June 8 in $J$. We do not detect any source
at the location of \sn\ and derive the following limiting AB
magnitudes: $J>23.5$ and $K_s>23.1$ (see
Table~\ref{tab:logphotometry}).

Using NOT images of \sn\ around peak magnitude, we project the
position of the supernova onto the sum of the 2013 September
and 2014 July Keck $g$-band images, shown by
the crossbars in Figure~\ref{fig:hostimages}. The uncertainty in this
projection is 0.4 LRIS pixels (0.05\arcsec). There does not appear to
be a significant offset between the location of the SN and the
peak of the galaxy light. There is an additional
object about 3\arcsec\ to the northeast of the \sn\ location, which
might be related to the host galaxy of \sn, but given its faintness
and the relatively large offset it is more likely to be unrelated. 

No emission lines are detected at the redshift of \sn\ in any of our
spectra. We determined the most stringent limits on the
[\ion{O}{2}]~\l 3727, [\ion{O}{3}]~\l 5007, and \ha\ observed line
fluxes to be $1.5\times10^{-18}$~\ergscm,
$1.5\times10^{-18}$~\ergscm, and $4.6\times10^{-18}$~\ergscm,
respectively.  Adopting the apparent host-galaxy $g$-band magnitude of
$g=27.0$ as the continuum flux (corresponding to
$8.8\times10^{-20}$~\ergscmA), these flux limits correspond to the
following rest-frame equivalent-width limits: $W_r ({\rm
  [O~II]~\lambda 3727})<10$~\AA, $W_r ({\rm [O~III]~\lambda
  5007})<10$~\AA, and $W_r ({\rm H\alpha})<30$~\AA.

We convert the [\ion{O}{2}] and \ha\ flux limits to star-formation
rate (SFR) limits of SFR$_{\rm [O~II]}<0.07~\Msunyr$ and SFR$_{\rm
  H\alpha}< 0.15~\Msunyr$, using the [\ion{O}{2}]-SFR conversion
derived by \citet{Savaglio09} for low-luminosity star-forming GRB host
galaxies at $z<0.5$ and the \ha-SFR conversion provided by
\citet{1998ARA&A..36..189K}. In the [\ion{O}{2}] flux-to-SFR
conversion, we have conservatively increased the limit by a factor of
3.4, which is twice the dispersion of the [\ion{O}{2}]-SFR conversion
factor \citep[see][]{Savaglio09}. In the \ha\ conversion we use the
initial mass function (IMF) of \citet{2003ApJ...593..258B}; this
results in an SFR that is a factor of 1.8 lower than when adopting a
Salpeter IMF.  We note that we have not included any dust or slit-loss
corrections in these SFR limits.  As an example, an extinction of
$A_V=0.5$~mag would increase the SFR$_{\rm [O~II]}$ limit by a factor
of about two. From the lack of absorption at the expected wavelengths
of the \ion{Na}{1}~D doublet, we can constrain the host-galaxy
extinction. The upper limit on the equivalent width of \ion{Na}{1}~D2
of $W_{\rm rest}<0.5$~\AA\ (the spectrum S/N around the expected
wavelength of this line is 2.5-3) provides $E_{B-V}<0.12$~mag
\citep{2012MNRAS.426.1465P}, or $A_V \lesssim 0.4$~mag.  Given the
very small amount of neutral metal absorption, the extinction along
the \sn\ sight line and in the host galaxy is likely smaller than the
limit inferred from \ion{Na}{1}~D: $A_V\lesssim 0.1$~mag.

We searched for potential diffuse interstellar bands (DIBs),
but do not detect any. DIBs have been successfully detected in the
host galaxies of SN~2001el and SN~2003hn \citep{2005A&A...429..559S}
and have been observed to vary with time in spectra of the broad-lined
Type Ic supernova SN 2012ap \citep{2014ApJ...782L...5M}. The
S/N of the X-shooter spectra in the near-infrared
region, where the \sn\ DIBs would be located, is so low (S/N $\approx 3$)
that we cannot put meaningful constraints on their strength.

\section{Discussion}
\label{sec:discussion}

Gas clouds similar to the one in the host of \sn\ are observed in
halos of high-redshift galaxies. \ion{Mg}{2} absorbers are routinely
detected in QSO spectra
\citep[e.g.,][]{2005ApJ...628..637N,2011MNRAS.416.3118K}. QSO
\ion{Mg}{2} absorption equivalent widths at $0.07 < z < 1.1$ as a
function of galaxy impact parameter have recently been reported by
\citet{2013ApJ...776..114N}. About 60\% of the detections have
equivalent widths above the one detected in \sn:
$W_r(2796)=0.51\pm0.02$~\AA. This fraction increases to 74\% when
considering the absorbers within 20~kpc of their galaxy
counterpart. Compared to the sample of \ion{Mg}{2} absorbers with
impact parameters to their galaxy counterparts collected by
\citet{2010ApJ...714.1521C}, this latter number is very similar: 78\%
of absorbers within 20~kpc of a galaxy have a larger
\ion{Mg}{2}~\l2796 equivalent width than the \sn\ absorber.  The
\ion{Mg}{2}~\l2796 equivalent width of \sn\ is lower than all but one
of a sample of 23 sub-DLAs (absorbers along QSO sightlines with $19
\lesssim$ log $N$(\ion{H}{1})$_{\rm sub-DLA} < 20.3$) at $z<1.5$
\citep{2008MNRAS.384.1015M,2008MNRAS.386.2209P,2009MNRAS.393.1513M}. This
shows that the gas column density along the sightline to \sn\ is low
compared to a sample of random sightlines through galaxies in the
foreground of QSOs. Following \citet{2006MNRAS.368..335E}, we
calculate the $D$-index, which can help determine whether a
\ion{Mg}{2} absorber is a DLA (with log $N$(\ion{H}{1}) $\geq 20.3$),
when the \ion{H}{1} column density is not known. For \sn, $D = 1000
\times W_r(2796) / \Delta V = 6.7$ (with $\Delta V = 76$~\kms;
Sect.~\ref{sec:narrowabslines}); this value is higher than the
recommended $D$-index DLA threshold of 5.1 for our resolution, and
therefore the \sn\ absorber has a reasonable probability ($\sim 90$\%)
of being a DLA.

\citet{2013arXiv1311.0026L} have recently studied a sample of 31 host
galaxies of hydrogen-poor SLSNe, showing that SLSN-I hosts appear to
have some similarities to GRB hosts. We note that, although unbiased
samples of GRB host galaxies do exist
\citep[e.g.,][]{2012ApJ...756..187H}, the sample of GRB hosts that
Lunnan et al. use to compare with SLSNe \citep[mostly
  from][]{2010MNRAS.405...57S} is biased toward brighter GRB hosts.
For the SLSN-I sample, on the other hand, host galaxies that are not
detected are included in the analysis. Also, the most massive SLSN
host galaxy in the sample of \citet{2013arXiv1311.0026L} is that of
PS1-10afx, which was shown by \citet{2013ApJ...768L..20Q} to be a
gravitationally lensed SN~Ia.

An interesting alternative method of probing the environments of
SLSNe, thereby providing a potential way to constrain the nature of
the progenitor, is through absorption-line spectroscopy.  In this
paper we present the first \ion{Mg}{1}, \ion{Mg}{2}, and \ion{Fe}{2}
column-density measurements of a SLSN, thanks to the sensitivity and
intermediate resolving power of X-shooter.  Normally, spectra of
SLSNe are observed at low resolution, which allows only for the
measurement of the equivalent width of the absorption features.

\begin{deluxetable}{lllllll}
  \tablecaption{Absorption-line strengths in Type I SLSNe \label{tab:ews}}
  \tablehead{
    \colhead{ID} & 
    \colhead{$z$} &
    \multicolumn{3}{c}{$W_{\rm r}$ (\AA)} &
    \colhead{Ref.} \\
    &  
    & 
    \colhead{\ion{Fe}{2}~\l 2600} &
    \colhead{\ion{Mg}{2}~\l 2800} &
    \colhead{\ion{Mg}{1}~\l 2852} & 
  }
  \startdata
  SN~2005ap                  & 0.2832 & $<0.8$  & $2.92\pm0.16$ & $0.39\pm0.10$ & 1 \\
  SN~2006oz                  & 0.396  & $<3.0$  & $3.43\pm0.84$ & $<1.2$        & 2 \\
  SCP~06F6                   & 1.189  & $<0.5$  & $0.93\pm0.24$ & $<0.3$        & 6,4 \\
  PS1-10ky                   & 0.9558 & $<0.7$  & $1.87\pm0.19$ & \nodata       & 3 \\
  PS1-10awh                  & 0.9084 & $<1.0$  & $2.79\pm0.32$ & $<0.8$        & 3 \\
  PTF~09atu                  & 0.501  & \nodata & $2.21\pm0.16$ & $0.51\pm0.08$ & 4 \\
  PTF~09cnd                  & 0.258  & \nodata & $2.69\pm0.15$ & $0.33\pm0.09$ & 4 \\
  PTF~09cwl\tablenotemark{a} & 0.349  & \nodata & $1.02\pm0.13$ & $0.42\pm0.09$ & 4 \\
  PTF~10cwr\tablenotemark{b} & 0.230  & $<1.0$  & $3.53\pm0.24$ & $<0.45$       & 4 \\
  SNLS-06D4eu                & 1.588  & $<1.2$  & $2.78\pm0.51$ & $<1.3$        & 5 \\
  SNLS-07D2bv                & 1.50   & $<1.5$  & $4.40\pm0.59$ & $<1.8$        & 5 \\
  SN~2013dg\tablenotemark{c} & 0.26   & $<0.7$  & $<0.4$        & $<0.2$        & 7
  \enddata
  \tablecomments{References: [1] \citet{2007ApJ...668L..99Q}; [2]
    \citet{2012A&A...541A.129L}; [3] \citet{2011ApJ...743..114C}; [4]
    \citet{2011Natur.474..487Q}; [5] \citet{2013ApJ...779...98H}; 
          [6] \citet{2009ApJ...690.1358B}; [7] \citet{2014arXiv1405.1325N}.}
  \tablenotetext{a}{= SN~2009jh.}
  \tablenotetext{b}{= SN~2010gx = CSS100313 J112547-084941.}
  \tablenotetext{c}{= MLS130517:131841-070443 = CSS130530:131841-070443.}  
\end{deluxetable}

Using published spectra of hydrogen-poor SLSNe available through the
Weizmann interactive SN data repository
\citep[WISeREP;][]{2012PASP..124..668Y}, we have measured 
$W_{\rm r}$ of the \ion{Mg}{1}~\l 2852
and \ion{Mg}{2}~$\lambda\lambda$2796, 2803 absorption lines for about a dozen
SLSNe-I. We also attempted to measure the same quantity for the
\ion{Fe}{2} \l2600 absorption line, but the low redshift and/or low
S/N of the spectra in the blue results in the
nondetection of this line in most cases.  Table~\ref{tab:ews} lists the 
SNe with publicly available spectra for which the SN redshift is sufficiently
high ($z\gtrsim0.2$) for the relevant transitions to shift into the
wavelength range covered by the spectra.  We report upper limits in
cases where the line is covered by the spectrum but not detected. Most
spectra lack an associated error spectrum; we therefore estimate the
flux error per pixel by measuring the scatter in the continuum free
from absorption lines and close to the relevant transition. The
normalized flux error, the number of pixels used in the equivalent
width measurement, and the pixel size in \AA\ determine the 
uncertainty in the equivalent width.

\begin{figure}
  \centering
  \includegraphics[width=0.6\hsize]{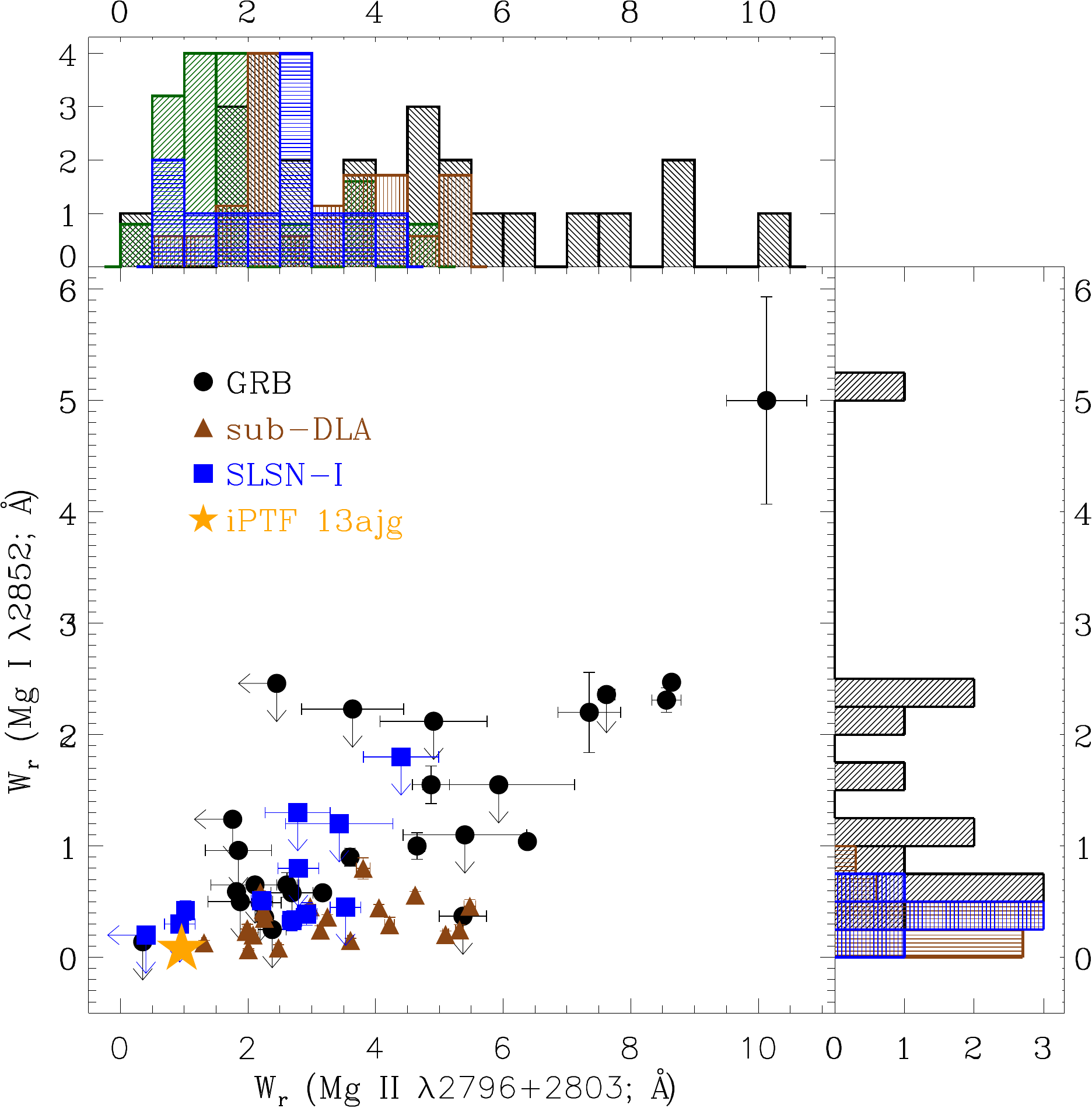}
  \caption{Comparison of the rest-frame \ion{Mg}{1} \l 2852 and
    \ion{Mg}{2} $\lambda\lambda$2796, 2803 equivalent widths ($W_{\rm
      r}$) measured in spectra of GRB afterglows \citep[black
      circles;][]{2012A&A...548A..11D}, sub-DLAs in the foreground of
    background QSOs \citep[brown
      triangles;][]{2008MNRAS.384.1015M,2008MNRAS.386.2209P,2009MNRAS.393.1513M},
    and hydrogen-poor SLSNe with published spectra (blue squares; see
    Table~\ref{tab:ews} for details). The average strength of the
    \ion{Mg}{1} and \ion{Mg}{2} absorption in SLSN host galaxies is
    significantly lower than in GRB hosts, and is similar to that of
    sub-DLAs. The measurements for \sn\ are indicated with the orange
    star. In the \ion{Mg}{2} histogram, we also plot (in green) a
    sample of \ion{Mg}{2} absorbers along QSO sightlines that are
    within 20~kpc of a nearby galaxy \citep{2013ApJ...776..114N}. The
    histograms (which do not include the upper limits) of the
    \ion{Mg}{2} absorbers and sub-DLAs have been normalized to match
    the peak of the SLSN histograms. \label{fig:mg1mg2}}
\end{figure}

In Figure~\ref{fig:mg1mg2}, we compare the $W_r$ of \ion{Mg}{1} and
\ion{Mg}{2} (sum of $\lambda$2796 and $\lambda$2803) of a sample of
hydrogen-poor SLSNe with those measured for GRBs, taken from
\citet{2012A&A...548A..11D}, and a sample of sub-DLAs in the
foreground of background QSOs
\citep{2008MNRAS.384.1015M,2008MNRAS.386.2209P,2009MNRAS.393.1513M}.
The GRB sample includes a total of 69 afterglows in the redshift range
$0.12<z<6.7$ with low-resolution optical spectroscopy available. We
include all GRBs (28) for which \citet{2012A&A...548A..11D} list
detections and/or upper limits for both the \ion{Mg}{1} and
\ion{Mg}{2} host-galaxy absorption lines. The redshift range for this
subsample is $0.4<z<2.4$ with a median value of $z=1.2$. The sub-DLA
sample redshift range is $0.68<z<1.55$ with a median of $z=0.95$. For
the SLSN-I sample, the redshift range is $0.2<z<1.6$ with a median of
$z=0.5$.

On average, the absorption-line strengths in the environments of SLSNe
are lower than those measured for GRB host galaxies. Applying a
Kolmogorov-Smirnov (KS) test \citep[see][]{numrec}, we find a
probability that the GRB and SLSN \ion{Mg}{1} and \ion{Mg}{2}
distributions are taken from the same parent population to be 0.16\%
and 2.3\%, respectively. Performing the same KS tests when compared to
the sub-DLA sample, the corresponding \ion{Mg}{1} and \ion{Mg}{2}
probabilities are 61\% and 27\%, respectively. This suggests that the
environments of SLSNe and GRBs are different
\citep[see][]{2013ApJ...779..114V}. We note that this inference would
not be in disagreement with SLSNe and GRBs exploding in galaxies that
appear to be similar, as argued by \citet{2013arXiv1311.0026L}. 

It is important to note that even though the difference in absorption
strengths in SLSN and GRB hosts appears to be significant, it might be
that the sample of SLSNe is biased toward sightlines having low
extinction and hence low gas column densities. GRB afterglows in
principle suffer from the same bias, and they are detected at higher
redshifts, leading to the observed extinction to be higher, on
average, for GRB sightlines than for SLSNe. But since GRBs are also
much brighter at early times, only the dustier sightlines are being
missed in the GRB case. For SLSNe the bias could be significant, also
because the redshift needs to be $z\gtrsim0.2$ for the Mg absorption
lines to shift into the optical range and for the SN to make it onto
Figure~\ref{fig:mg1mg2}; this redshift requirement results in
apparently fainter SNe, which are more difficult to detect.  On the
other hand, several of the targets listed in Table~\ref{tab:ews} have
peak magnitudes well above the limiting magnitude of the survey that
discovered them, which suggests that this bias is probably not too
important.

In summary, a strong bias is not likely to be present, and
Figure~\ref{fig:mg1mg2} provides a strong indication that the
environments of SLSNe-I and GRBs are different. In any case, expanding
the number of SLSNe with measurements of the strength of narrow
low-ionization ISM absorption lines, using either equivalent widths
but preferably column densities, can provide useful clues to the place
of birth and progenitor nature of SLSNe \citep[see
  also][]{2012ApJ...755L..29B}.

Different models have been invoked for explaining the energetics
observed in hydrogen-poor SLSNe, such as a pair-instability SN
\citep{1967PhRvL..18..379B,2009Natur.462..624G}, the interaction of
the SN ejecta with a dense circumstellar medium
\citep{2011ApJ...729L...6C,2010arXiv1009.4353B,2013ApJ...773...76C},
additional energy injection by a highly magnetized millisecond pulsar
or magnetar \citep{2010ApJ...717..245K, 2010ApJ...719L.204W}, or
fall-back accretion \citep{2013ApJ...772...30D}. The late-time decay
of \sn\ is too rapid for the light curve to be powered by radioactive
decay of $^{56}$Ni, and therefore the pair-instability SN is
not compatible with the observations of \sn.  In the interaction
model, the circumstellar medium is required to be hydrogen-poor given the
lack of narrow hydrogen emission lines in the spectra of \sn; such a
medium could potentially have been created by a pulsational 
pair-instability SN
\citep{2007Natur.450..390W,2012ApJ...760..154C,2014ApJ...785...37B}. 

Interestingly, the less energetic cousins of the hydrogen-rich (Type
II) superluminous SNe, the SNe~IIn, are usually found in regions of
their host galaxies with little recent star formation as traced by
\ha\ \citep{2014MNRAS.441.2230H}. Such a location distribution would
be compatible with the absorption-line strengths that we measured for
a sample of H-poor SLSNe. Although the bolometric light curve of
\sn\ is consistent with additional energy input from a magnetar with a
magnetic field strength of $B=2.3\times10^{14}$~G and an initial spin
period of $P=1.1$~ms (Sect.~\ref{sec:Lbol}), similar to what has been
found for other hydrogen-poor SLSNe
\citep[e.g.,][]{2013ApJ...770..128I}, given the generic nature and
flexibility of the magnetar model it would be premature to conclude
that it accounts for the energetics of \sn. Perhaps owing to their
generic nature, magnetars have also been invoked as a possible central
engine of GRBs
\citep[e.g.,][]{1992Natur.357..472U,2001ApJ...552L..35Z,2009MNRAS.396.2038B}.

Placing \sn\ at higher redshifts, assuming our best estimate for the
flux evolution as a function of time and wavelength
(Sect.~\ref{sec:Lbol}), it would have an apparent peak magnitude of
$z_{\rm AB}=22.5$ and $24.5$ at redshifts of 2 and 3, respectively.
Therefore, the next generation of 30-m class telescopes, such as the
European Extremely Large Telescope (E-ELT) and the Thirty Meter
Telescope (TMT), will be capable of studying in detail the host-galaxy
ISM of superluminous explosions at an age of the Universe when the
star-formation activity was peaking.  These redshifts will also allow
the detection of \lya\ absorption from the ground, making it possible
to measure \ion{H}{1} column densities and metallicities.

\section{Conclusions}
\label{sec:conclusions}

We have presented an extensive imaging and spectroscopic campaign on a
superluminous supernova discovered by the intermediate Palomar
Transient Factory: \sn. This is one of the most luminous SNe to
date, with an absolute magnitude at peak of $M_{u,{\rm AB}}=-22.5$
($M_{U,{\rm Vega}}=-23.5$, $M_{B,{\rm Vega}}=-22.2$ mag). 

We infer the photospheric radius and temperature evolution as a
function of time using the nine epochs of Keck and VLT spectra
spanning the phases from $-9$ to $+80$ days, from which we estimate
the bolometric light curve.  The observed bolometric peak luminosity
of \sn\ is $3.2\times 10^{44}$~erg~\persec, while the estimated total
radiated energy is $1.3\times10^{51}$~erg. The bolometric light curve
is consistent with a highly magnetized, rapidly rotating neutron star
spinning down and providing its rotational energy to boost the
supernova energetics. However, owing to the generic nature and
flexibility of the magnetar model, it would be premature to conclude
that it is accounting for the energetics of \sn.

Intermediate-resolution spectra ($R\approx6000$) of \sn\ obtained with
X-shooter enabled us to infer the metal column densities of the
following UV absorption-line features: log
$N$(\ion{Mg}{1}) $=11.94\pm0.06$, log $N$(\ion{Mg}{2}) $=14.7\pm0.3$, and
log $N$(\ion{Fe}{2}) $=14.25\pm0.10$.  These column densities, as well
as \ion{Mg}{1} and \ion{Mg}{2} equivalent widths of a sample of
hydrogen-poor SLSNe taken from the literature, are at the low end of
those derived for GRBs, also sites of massive-star
formation. This suggests that the environments in which SLSNe and GRBs
explode are different. From the nondetection of \ion{Fe}{2}
fine-structure absorption lines, we derive a strict lower limit on the
distance between the SN and the narrow-line absorbing gas of
50~pc. The velocity width of the features is narrow, $\Delta
V=76$~\kms, indicating a low-mass host galaxy with an estimated
metallicity of [M/H] $\lesssim -0.7$.

No host-galaxy emission lines are detected, leading to an upper limit
on the star-formation rate of SFR$_{\rm
  [O~II]}<0.07~\Msunyr$. Late-time imaging shows the host galaxy of
\sn\ to be faint, with $g_{\rm AB} \approx 27.0$ and $R_{\rm AB} \geq
26.0$ mag.

\begin{acknowledgements}
  
  This paper is based on observations collected at the Palomar 48 and
  60~inch telescope, the Nordic Optical Telescope (NOT), the Discovery Channel
  Telescope (DCT), the Very Large Telescope (VLT) under proposal
  no. 291.D-5009, and the Keck-I and Keck-II telescopes. The W. M. Keck
  Observatory is
  operated as a scientific partnership among the California Institute
  of Technology, the University of California, and the National
  Aeronautics and Space Administration (NASA); it was made
  possible by the generous financial support of the W. M. Keck
  Foundation. We are grateful to Ori D. Fox, Isaac Shivvers, Patrick
  L. Kelly, WeiKang Zheng, Sumin Tang, W. Kao, and Joel Johansson for
  performing part of the Keck observations presented in this paper,
  and to ESO's User Support Department and Paranal observing staff
  for arranging and securing the VLT DDT observations. These results
  made use of Lowell Observatory's Discovery Channel Telescope; 
  Lowell operates the DCT in partnership with Boston University,
  Northern Arizona University, the University of Maryland, and the
  University of Toledo. Partial support of the DCT was provided by
  Discovery Communications. The LMI at the DCT was built by Lowell 
  Observatory using funds from the NSF grant AST--1005313. We wish to
  thank Sylvain Veilleux, Antonino Cucchiara, Suvi Gezari, and
  Eleonora Troja for assistance in obtaining the DCT data. It is a
  pleasure to thank Daniele Malesani for providing his handy finder-chart 
  routine, and Melina Bersten for interesting discussions.
  A.G.-Y. is supported by the EU/FP7 via ERC grant No. 307260, the
  Quantum Universe I-Core program by the Israeli Committee for
  planning and funding, and the ISF, GIF, Minerva, and ISF grants,
  WIS-UK ``making connections,'' and Kimmel and ARCHES awards.
  A.V.F.'s supernova group at UC Berkeley is supported through NSF grant
  AST--1211916, the TABASGO Foundation, and the Christopher R. Redlich Fund.
  M.S. acknowledges support from the Royal Society. Support for D.A.P. was
  provided by NASA through Hubble Fellowship grant HST-HF-51296.01-A
  awarded by the Space Telescope Science Institute, which is operated
  by the Association of Universities for Research in Astronomy, Inc.,
  for NASA, under contract NAS 5-26555. The Dark Cosmology Centre is
  funded by the DNRF.  The National Energy Research Scientific
  Computing Center, which is supported by the Office of Science of the
  U.S. Department of Energy under Contract No. DE-AC02-05CH11231,
  provided staff, computational resources, and data storage for this
  project.

\end{acknowledgements}

\bibliographystyle{apj} 
\bibliography{references}

\appendix
\setcounter{table}{0}

\begin{deluxetable}{crrrrrcl}
  \tablecaption{Log of imaging observations of \sn\label{tab:logphotometry}}
  \tablehead{
    \colhead{MJD} & 
    \colhead{JD-JD$_{\rm Rpeak}$\tablenotemark{a}} &
    \colhead{Phase} &
    \colhead{Telescope} & 
    \colhead{Exp.~Time} & 
    \colhead{Filter} & 
    \colhead{Seeing} &
    \colhead{Magnitude\tablenotemark{b}} \\
    (days) & (days) & (days) & & (min) & & \arcsec & AB
  }
  \startdata
    56324.553  & -81.047 & -46.571 &        P48  & $2 \times 1.0$  &   $R$  &  2.5  & $>$ 21.43 \\
    56327.550  & -78.050 & -44.849 &        P48  &   1.0  &   $R$  &  2.5  & $>$ 21.50 \\
    56341.043  & -64.557 & -37.095 &        P48  & $2 \times 1.0$  &   $R$  &  2.6  & $>$ 22.00 \\
    56356.484  & -49.116 & -28.223 &        P48  &   1.0  &   $R$  &  2.7  & $>$ 20.87 \\
    56362.969  & -42.631 & -24.496 &        P48  & $6 \times 1.0$  &   $R$  &  2.8  &     21.26 $\pm$ 0.14 \\
    56365.545  & -40.055 & -23.016 &        P48  & $3 \times 1.0$  &   $R$  &  2.5  &     21.67 $\pm$ 0.43 \\
    56367.502  & -38.098 & -21.892 &        P48  & $2 \times 1.0$  &   $R$  &  2.7  & $>$ 20.31 \\
    56369.448  & -36.152 & -20.773 &        P48  & $3 \times 1.0$  &   $R$  &  2.8  &     21.24 $\pm$ 0.15 \\
    56372.390  & -33.210 & -19.083 &        P48  & $3 \times 1.0$  &   $R$  &  2.7  & $>$ 19.89 \\
    56374.945  & -30.655 & -17.615 &        P48  & $6 \times 1.0$  &   $R$  &  2.7  &     20.91 $\pm$ 0.12 \\
    56376.452  & -29.148 & -16.749 &        P48  & $3 \times 1.0$  &   $R$  &  2.2  &     20.75 $\pm$ 0.15 \\
    56382.382  & -23.218 & -13.341 &        P48  &   1.0  &   $R$  &  2.8  &     20.70 $\pm$ 0.30 \\
    56385.884  & -19.716 & -11.329 &        P48  & $6 \times 1.0$  &   $R$  &  2.7  &     20.64 $\pm$ 0.10 \\
    56387.956  & -17.644 & -10.138 &        P48  & $4 \times 1.0$  &   $R$  &  2.9  &     20.58 $\pm$ 0.10 \\
    56389.389  & -16.211 &  -9.315 &        P48  & $3 \times 1.0$  &   $R$  &  2.6  &     20.59 $\pm$ 0.09 \\
    56394.993  & -10.607 &  -6.095 &        P48  & $3 \times 1.0$  &   $R$  &  2.6  &     20.34 $\pm$ 0.08 \\
    56396.325  &  -9.275 &  -5.330 &        P48  & $3 \times 1.0$  &   $R$  &  3.0  &     20.30 $\pm$ 0.09 \\
    56399.891  &  -5.709 &  -3.280 &        P48  & $6 \times 1.0$  &   $R$  &  2.9  &     20.44 $\pm$ 0.07 \\
    56401.899  &  -3.701 &  -2.127 &        P48  & $6 \times 1.0$  &   $R$  &  2.9  &     20.26 $\pm$ 0.07 \\
    56403.882  &  -1.718 &  -0.987 &        P48  & $6 \times 1.0$  &   $R$  &  2.2  &     20.23 $\pm$ 0.06 \\
    56405.397  &  -0.203 &  -0.117 &        P48  & $3 \times 1.0$  &   $R$  &  2.5  &     20.14 $\pm$ 0.13 \\
    56412.775  &   7.175 &   4.123 &        P48  & $5 \times 1.0$  &   $R$  &  2.7  &     20.39 $\pm$ 0.10 \\
    56415.852  &  10.252 &   5.891 &        P48  & $6 \times 1.0$  &   $R$  &  2.4  &     20.34 $\pm$ 0.07 \\
    56417.306  &  11.706 &   6.726 &        P48  & $2 \times 1.0$  &   $R$  &  3.1  &     20.32 $\pm$ 0.10 \\
    56422.881  &  17.281 &   9.930 &        P48  & $6 \times 1.0$  &   $R$  &  2.5  &     20.42 $\pm$ 0.06 \\
    56424.839  &  19.239 &  11.055 &        P48  & $6 \times 1.0$  &   $R$  &  2.5  &     20.46 $\pm$ 0.06 \\
    56426.743  &  21.143 &  12.149 &        P48  & $5 \times 1.0$  &   $R$  &  2.7  &     20.47 $\pm$ 0.07 \\
    56428.571  &  22.971 &  13.199 &        P48  & $4 \times 1.0$  &   $R$  &  2.7  &     20.57 $\pm$ 0.11 \\
    56431.886  &  26.286 &  15.104 &        P48  & $6 \times 1.0$  &   $R$  &  2.5  &     20.68 $\pm$ 0.09 \\
    56433.845  &  28.245 &  16.230 &        P48  & $6 \times 1.0$  &   $R$  &  2.5  &     20.56 $\pm$ 0.10 \\
    56440.358  &  34.758 &  19.972 &        P48  & $2 \times 1.0$  &   $R$  &  3.0  &     20.58 $\pm$ 0.29 \\
    56442.902  &  37.302 &  21.434 &        P48  & $4 \times 1.0$  &   $R$  &  2.5  &     21.13 $\pm$ 0.15 \\
    56444.810  &  39.210 &  22.531 &        P48  & $4 \times 1.0$  &   $R$  &  2.3  &     21.26 $\pm$ 0.14 \\
    56446.752  &  41.152 &  23.646 &        P48  & $4 \times 1.0$  &   $R$  &  2.1  &     21.22 $\pm$ 0.12 \\
    56448.747  &  43.147 &  24.793 &        P48  & $4 \times 1.0$  &   $R$  &  2.2  &     21.41 $\pm$ 0.14 \\
    56450.748  &  45.148 &  25.943 &        P48  & $4 \times 1.0$  &   $R$  &  2.3  &     21.04 $\pm$ 0.11 \\
    56452.744  &  47.144 &  27.090 &        P48  & $4 \times 1.0$  &   $R$  &  2.5  &     21.27 $\pm$ 0.13 \\
    56454.739  &  49.139 &  28.236 &        P48  & $4 \times 1.0$  &   $R$  &  2.3  &     21.39 $\pm$ 0.15 \\
    56456.738  &  51.138 &  29.385 &        P48  & $4 \times 1.0$  &   $R$  &  2.3  &     21.60 $\pm$ 0.18 \\
    56458.735  &  53.135 &  30.532 &        P48  & $4 \times 1.0$  &   $R$  &  2.2  &     21.55 $\pm$ 0.18 \\
    56460.730  &  55.130 &  31.678 &        P48  & $4 \times 1.0$  &   $R$  &  2.3  &     21.33 $\pm$ 0.18 \\
    56462.727  &  57.127 &  32.826 &        P48  & $4 \times 1.0$  &   $R$  &  2.4  & $>$ 21.93 \\
    56469.205  &  63.605 &  36.548 &        P48  & $2 \times 1.0$  &   $R$  &  2.1  &     21.43 $\pm$ 0.23 \\
    56471.718  &  66.118 &  37.992 &        P48  & $4 \times 1.0$  &   $R$  &  2.4  &     21.66 $\pm$ 0.21 \\
    56473.718  &  68.118 &  39.142 &        P48  & $4 \times 1.0$  &   $R$  &  2.2  &     21.62 $\pm$ 0.26 \\
    56475.805  &  70.205 &  40.341 &        P48  & $4 \times 1.0$  &   $R$  &  2.6  & $>$ 21.68 \\
    56477.297  &  71.697 &  41.198 &        P48  & $2 \times 1.0$  &   $R$  &  2.2  &     21.68 $\pm$ 0.28 \\
    56486.725  &  81.125 &  46.616 &        P48  & $4 \times 1.0$  &   $R$  &  2.1  &     22.21 $\pm$ 0.30 \\
    56488.800  &  83.200 &  47.808 &        P48  & $4 \times 1.0$  &   $R$  &  2.3  & $>$ 22.26 \\
    56490.904  &  85.304 &  49.017 &        P48  & $3 \times 1.0$  &   $R$  &  2.7  & $>$ 20.97 \\
  &&&&&&& \\
    56395.324  & -10.276 &  -5.905 &        P60  &   2.0  &   $i$  &  1.6  &     20.44 $\pm$ 0.09 \\
    56395.326  & -10.274 &  -5.904 &        P60  &   2.0  &   $r$  &  1.5  &     20.39 $\pm$ 0.05 \\
    56395.327  & -10.273 &  -5.903 &        P60  &   2.0  &   $B$  &  1.4  &     20.92 $\pm$ 0.10 \\
    56395.329  & -10.271 &  -5.902 &        P60  &   2.0  &   $g$  &  1.4  &     20.74 $\pm$ 0.06 \\
    56406.208  &   0.607 &   0.349 &        P60  &   3.0  &   $i$  &  1.6  &     20.38 $\pm$ 0.22 \\
    56407.206  &   1.606 &   0.923 &        P60  & $2 \times 3.0$  &   $B$  &  1.7  & $>$ 20.62 \\
    56408.200  &   2.600 &   1.494 &        P60  &   3.0  &   $i$  &  1.8  &     20.38 $\pm$ 0.28 \\
    56412.442  &   6.842 &   3.932 &        P60  &   3.0  &   $i$  &  1.8  &     20.41 $\pm$ 0.09 \\
    56412.444  &   6.844 &   3.933 &        P60  &   3.0  &   $r$  &  1.6  &     20.23 $\pm$ 0.06 \\
    56412.449  &   6.849 &   3.936 &        P60  &   3.0  &   $g$  &  1.8  &     20.82 $\pm$ 0.10 \\
    56413.913  &   8.313 &   4.777 &        P60  & $4 \times 3.0$  &   $B$  &  2.1  &     21.32 $\pm$ 0.12 \\
    56416.323  &  10.723 &   6.162 &        P60  &   3.0  &   $B$  &  1.1  &     21.21 $\pm$ 0.09 \\
    56422.412  &  16.812 &   9.660 &        P60  &   3.0  &   $i$  &  1.5  &     20.30 $\pm$ 0.07 \\
    56422.415  &  16.815 &   9.662 &        P60  &   3.0  &   $B$  &  1.6  &     21.47 $\pm$ 0.14 \\
    56423.368  &  17.768 &  10.210 &        P60  &   3.0  &   $r$  &  1.6  &     20.40 $\pm$ 0.05 \\
    56423.371  &  17.771 &  10.211 &        P60  &   3.0  &   $g$  &  1.9  &     20.94 $\pm$ 0.06 \\
    56427.339  &  21.739 &  12.492 &        P60  &   3.0  &   $i$  &  2.7  &     20.63 $\pm$ 0.12 \\
    56428.358  &  22.758 &  13.077 &        P60  & $2 \times 3.0$  &   $B$  &  2.2  &     21.69 $\pm$ 0.28 \\
    56429.372  &  23.772 &  13.660 &        P60  &   3.0  &   $r$  &  2.3  &     19.26 $\pm$ 0.40 \\
    56429.377  &  23.777 &  13.663 &        P60  &   3.0  &   $g$  &  2.5  & $>$ 20.19 \\
    56431.376  &  25.776 &  14.811 &        P60  &   3.0  &   $r$  &  2.3  &     20.53 $\pm$ 0.07 \\
    56431.380  &  25.780 &  14.814 &        P60  &   3.0  &   $g$  &  3.1  &     21.56 $\pm$ 0.16\\
    56432.309  &  26.709 &  15.347 &        P60  &   3.0  &   $r$  &  1.3  &     20.68 $\pm$ 0.06 \\
    56432.314  &  26.714 &  15.350 &        P60  &   3.0  &   $g$  &  1.4  &     21.26 $\pm$ 0.11 \\
    56432.825  &  27.225 &  15.644 &        P60  & $4 \times 3.0$  &   $B$  &  1.6  &     21.76 $\pm$ 0.23 \\
    56433.311  &  27.711 &  15.923 &        P60  &   3.0  &   $i$  &  1.4  &     20.60 $\pm$ 0.11 \\
    56433.313  &  27.713 &  15.924 &        P60  &   3.0  &   $r$  &  1.2  &     20.83 $\pm$ 0.12 \\
    56433.317  &  27.717 &  15.927 &        P60  &   3.0  &   $g$  &  1.3  &     21.38 $\pm$ 0.16 \\
    56436.336  &  30.736 &  17.661 &        P60  & $3 \times 3.0$  &   $B$  &  2.4  & $>$ 21.05 \\
    56440.339  &  34.739 &  19.962 &        P60  &   3.0  &   $i$  &  2.2  &     20.70 $\pm$ 0.17 \\
    56440.341  &  34.741 &  19.963 &        P60  &   3.0  &   $r$  &  2.2  &     21.18 $\pm$ 0.22 \\
    56440.346  &  34.746 &  19.966 &        P60  &   3.0  &   $g$  &  2.1  &     22.77 $\pm$ 0.85 \\
    56441.363  &  35.763 &  20.550 &        P60  &   3.0  &   $g$  &  3.0  & $>$ 21.76 \\
    56441.639  &  36.039 &  20.708 &        P60  & $3 \times 3.0$  &   $B$  &  2.6  & $>$ 21.89 \\
    56443.209  &  37.609 &  21.611 &        P60  &   3.0  &   $i$  &  2.9  &     20.89 $\pm$ 0.17 \\
    56443.211  &  37.611 &  21.612 &        P60  &   3.0  &   $r$  &  2.7  &     21.11 $\pm$ 0.13 \\
    56443.216  &  37.616 &  21.615 &        P60  &   3.0  &   $g$  &  2.3  &     22.34 $\pm$ 0.22 \\
    56444.301  &  38.701 &  22.238 &        P60  &   3.0  &   $i$  &  2.6  &     21.06 $\pm$ 0.18 \\
    56444.306  &  38.706 &  22.241 &        P60  &   3.0  &   $g$  &  2.9  &     22.04 $\pm$ 0.20 \\
    56445.197  &  39.597 &  22.753 &        P60  &   3.0  &   $i$  &  1.2  &     21.01 $\pm$ 0.11 \\
    56445.202  &  39.602 &  22.756 &        P60  &   3.0  &   $g$  &  1.2  &     22.21 $\pm$ 0.14 \\
    56445.261  &  39.661 &  22.790 &        P60  & $3 \times 3.0$  &   $B$  &  2.0  &     22.38 $\pm$ 0.24 \\
    56447.834  &  42.234 &  24.268 &        P60  & $2 \times 3.0$  &   $B$  &  2.7  & $>$ 22.81 \\
    56449.294  &  43.694 &  25.107 &        P60  & $2 \times 3.0$  &   $r$  &  1.4  &     21.34 $\pm$ 0.06 \\
    56451.331  &  45.731 &  26.278 &        P60  &   3.0  &   $i$  &  1.7  &     21.04 $\pm$ 0.11 \\
    56451.825  &  46.225 &  26.562 &        P60  & $4 \times 3.0$  &   $B$  &  1.7  &     23.15 $\pm$ 0.31 \\
    56455.444  &  49.844 &  28.641 &        P60  &   3.0  &   $r$  &  3.2  &     21.57 $\pm$ 0.22 \\
    56456.216  &  50.616 &  29.085 &        P60  &   3.0  &   $r$  &  1.3  &     21.58 $\pm$ 0.10 \\
    56456.770  &  51.170 &  29.403 &        P60  & $4 \times 3.0$  &   $B$  &  1.6  &     23.10 $\pm$ 0.36 \\
    56457.207  &  51.607 &  29.654 &        P60  &   3.0  &   $i$  &  1.6  &     21.18 $\pm$ 0.14 \\
    56460.243  &  54.643 &  31.399 &        P60  & $3 \times 3.0$  &   $B$  &  1.9  & $>$ 22.53 \\
    56462.293  &  56.693 &  32.577 &        P60  &   3.0  &   $r$  &  3.2  & $>$ 21.78 \\
    56462.296  &  56.696 &  32.578 &        P60  &   3.0  &   $B$  &  3.1  & $>$ 21.30 \\
    56467.316  &  61.716 &  35.463 &        P60  &   3.0  &   $i$  &  2.0  &     20.97 $\pm$ 0.29 \\
    56467.319  &  61.719 &  35.465 &        P60  &   3.0  &   $r$  &  2.2  & $>$ 21.35 \\
    56468.446  &  62.846 &  36.112 &        P60  &   3.0  &   $i$  &  2.5  &     21.22 $\pm$ 0.32 \\
    56468.448  &  62.848 &  36.113 &        P60  &   3.0  &   $r$  &  2.2  & $>$ 21.46 \\
    56468.834  &  63.234 &  36.335 &        P60  & $4 \times 3.0$  &   $B$  &  1.6  & $>$ 21.88 \\
    56469.302  &  63.702 &  36.604 &        P60  &   3.0  &   $i$  &  2.0  &     21.53 $\pm$ 0.29 \\
    56469.305  &  63.705 &  36.606 &        P60  &   3.0  &   $r$  &  1.7  &     22.17 $\pm$ 0.35 \\
    56470.257  &  64.657 &  37.153 &        P60  &   3.0  &   $r$  &  1.3  &     21.61 $\pm$ 0.13 \\
    56473.303  &  67.703 &  38.903 &        P60  &   3.0  &   $i$  &  1.3  &     21.38 $\pm$ 0.16 \\
    56475.213  &  69.613 &  40.001 &        P60  &   3.0  &   $i$  &  1.9  &     21.21 $\pm$ 0.18 \\
    56477.303  &  71.703 &  41.202 &        P60  &   3.0  &   $i$  &  3.4  &     21.97 $\pm$ 0.49 \\
    56477.306  &  71.706 &  41.203 &        P60  &   3.0  &   $r$  &  2.6  & $>$ 21.47 \\
    56479.297  &  73.697 &  42.347 &        P60  &   3.0  &   $i$  &  1.8  &     21.65 $\pm$ 0.22 \\
    56479.299  &  73.699 &  42.348 &        P60  &   3.0  &   $r$  &  1.8  &     22.18 $\pm$ 0.25 \\
    56480.232  &  74.632 &  42.885 &        P60  &   3.0  &   $i$  &  1.4  &     21.67 $\pm$ 0.19 \\
    56480.235  &  74.635 &  42.886 &        P60  &   3.0  &   $r$  &  1.5  &     21.81 $\pm$ 0.14 \\
    &&&&&&& \\
    56418.126  &  12.526 &   7.198 & NOT+ALFOSC  & $3 \times 3.0$  &   $u$  &  0.6  &     22.46 $\pm$ 0.10 \\
    56418.133  &  12.533 &   7.202 & NOT+ALFOSC  & $3 \times 2.0$  &   $g$  &  0.8  &     20.89 $\pm$ 0.02 \\
    56418.139  &  12.539 &   7.205 & NOT+ALFOSC  & $3 \times 2.0$  &   $r$  &  0.7  &     20.37 $\pm$ 0.01 \\
    56418.144  &  12.544 &   7.208 & NOT+ALFOSC  & $3 \times 2.0$  &   $i$  &  0.7  &     20.45 $\pm$ 0.02 \\
    56418.151  &  12.551 &   7.212 & NOT+ALFOSC  & $3 \times 3.0$  &   $z$  &  0.8  &     20.55 $\pm$ 0.05 \\
    56430.077  &  24.477 &  14.065 & NOT+ALFOSC  & $4 \times 3.0$  &   $u$  &  0.7  &     22.58 $\pm$ 0.09 \\
    56430.084  &  24.484 &  14.069 & NOT+ALFOSC  & $3 \times 2.0$  &   $g$  &  0.7  &     21.27 $\pm$ 0.02 \\
    56430.089  &  24.489 &  14.072 & NOT+ALFOSC  & $3 \times 1.5$  &   $r$  &  0.7  &     20.63 $\pm$ 0.02 \\
    56430.094  &  24.494 &  14.075 & NOT+ALFOSC  & $3 \times 1.5$  &   $i$  &  0.7  &     20.53 $\pm$ 0.03 \\
    56430.100  &  24.500 &  14.078 & NOT+ALFOSC  & $3 \times 3.0$  &   $z$  &  0.6  &     20.64 $\pm$ 0.05 \\
    56477.038  &  71.438 &  41.049 & NOT+ALFOSC  & $5 \times 1.5$  &   $r$  &  0.6  &     22.08 $\pm$ 0.05 \\
    56477.046  &  71.446 &  41.054 & NOT+ALFOSC  & $5 \times 1.5$  &   $i$  &  0.6  &     21.73 $\pm$ 0.05 \\
    56477.057  &  71.457 &  41.060 & NOT+ALFOSC  & $5 \times 3.0$  &   $z$  &  0.6  &     21.68 $\pm$ 0.09 \\
    &&&&&&& \\
    56702.000  & 296.400 & 170.315 &    DCT+LMI  & $18 \times 3.0$ &   $r$  &  1.2  & $>$ 25.12 \\
    &&&&&&& \\
    56544.314  & 138.714 &  79.707 &  Keck+LRIS  & $2 \times 5.0$  &   $g$  &  0.6  &     26.89 $\pm$ 0.25 \\
    56544.314  & 138.714 &  79.707 &  Keck+LRIS  & $2 \times 5.0$  &  $R_s$  &  0.7  &     23.83 $\pm$ 0.08 \\
    56776.586  & 370.986 & 213.174 &  Keck+LRIS  & $4 \times 4.2$  &   $g$  &  0.6  &     26.49 $\pm$ 0.14 \\
    56776.586  & 370.986 & 213.174 &  Keck+LRIS  & $4 \times 4.0$  &  $R_s$  &  0.7  &     24.84 $\pm$ 0.09 \\
    56869.397  & 463.797 & 266.504 &  Keck+LRIS  & $3 \times 5.0$  &   $g$  &  0.8  & $>$ 26.9 \\
    56869.397  & 463.797 & 266.504 &  Keck+LRIS  & $3 \times 5.0$  &  $R_s$  &  0.9  &     26.01 $\pm$ 0.22 \\
    &&&&&&& \\
    56815.559  & 409.959 & 235.568 & Keck+MOSFIRE & $18 \times 23.3$  &  $K_s$ &  0.8  & $>$ 23.1 \\
    56816.533  & 410.933 & 236.128 & Keck+MOSFIRE &  $9 \times 30.6$  &  $J$  &  0.7  & $>$ 23.5 
  \enddata
  \tablenotetext{a}{JD$_{R,\rm peak} = 2,456,406.1.$}
  \tablenotetext{b}{The magnitudes have {\it not} been corrected for
    Galactic extinction.}
\end{deluxetable}

\end{document}